# Reviewing Evolution of Learning Functions and Semantic Information Measures for Understanding Deep Learning

**Chenguang Lu**

**Abstract:** A new trend in deep learning, represented by Mutual Information Neural Estimation (MINE) and Information Noise Contrast Estimation (InfoNCE), is emerging. In this trend, similarity functions and Estimated Mutual Information (EMI) are used as learning and objective functions. Coincidentally, EMI is essentially the same as Semantic Mutual Information (SeMI) proposed by the author 30 years ago. This paper first reviews the evolutionary histories of semantic information measures and learning functions. Then, it briefly introduces the author's semantic information G theory with the rate-fidelity function $R(G)$ ($G$ denotes SeMI, and $R(G)$ extends $R(D)$) and its applications to multi-label learning, the maximum Mutual Information (MI) classification, and mixture models. Then it discusses how we should understand the relationship between SeMI and Shannon's MI, two generalized entropies (fuzzy entropy and coverage entropy), Autoencoders, Gibbs distributions, and partition functions from the perspective of the $R(G)$ function or the G theory. An important conclusion is that mixture models and Restricted Boltzmann Machines converge because SeMI is maximized, and Shannon's MI is minimized, making information efficiency $G/R$ close to 1. A potential opportunity is to simplify deep learning by using Gaussian channel mixture models for pre-training deep neural networks' latent layers without considering gradients. It also discusses how the SeMI measure is used as the reward function (reflecting purposiveness) for reinforcement learning. The G theory helps interpret deep learning but is far from enough. Combining semantic information theory and deep learning will accelerate their development.

## 1. Introduction

Information-Theoretic Learning (ITL) has been used for a long time. The primary method is to use the likelihood function as the learning function, bring it into the cross-entropy, and then use the minimum cross-entropy or the minimum Kulback–Leibler (KL) divergence criterion to optimize model parameters. However, a new trend, represented by Mutual Information Neural Estimation (MINE) [1] and Information Noise Contrastive Estimation (InfoNCE) [2], is emerging in the field of deep learning. In this trend, researchers use similarity functions to construct parameterized mutual information (we call it Estimated Mutual Information (EMI)) that approximates to Shannon's Mutual Information (ShMI) [3], and then optimize the Deep Neural Network (DNN) by maximizing EMI. Abbreviation Section lists all abbreviations with original texts.

In January 2018, Belghazi et al. [1] published "MINE: Mutual Information Neural Estimation", which is the first article that uses the similarity function to construct EMI. In July 2018, Oord et al. [2] published "Representation Learning with Comparative Predictive Coding", proposing Information Noise Contrast Estimate (InfoNCE). This paper explicitly proposes that the similarity function is proportional to $P(x|y_j)/P(x)$ for expressing EMI. MINE and Inference achieve distinct successes and encourage others. In 2019,

Hjelm et al. [4,5] proposed Deep InfoMax (DIM) based on MINE. They combine DIM and InfoNCE (see Equation (5) in [4]) to achieve better results (they believe that InfoNCE was put forward independently of MINE). In 2020, Chen et al. proposed SimCLR [6], He et al. proposed MoCo [7], and Grill et al. presented BOYL [8], all of which show strong learning ability. All of them use similarity functions to construct the EMI or the loss function similar to that used for InfoNCE.

On the other hand, in response to the call from Weaver in 1949 [9], many researchers have been studying semantic information theory. The trend of researching semantic communication and semantic information theory is also gradually growing. One reason is the demand of next-generation Internet for semantic communication [10,11]. We need to use similarity or semantics instead of the Shannon channel to make probability predictions and data compression. Another reason is that semantic information theory has also made significant progress in machine learning. In particular, the author (of this paper) proposed a simple method of obtaining truth or similarity functions from sampling distributions in 2017 [12] and provided a group of channels matching algorithms for multi-label learning, the maximum mutual information classification of unseen instances, and mixture models [12,13].

Coincidentally, the EMI measure is essentially the same as the semantic information measure proposed by the author 30 years ago. In 1990, the author proposed using truth (or similarity) functions to construct semantic and estimated information measures [14]. In 1993, he published the monograph "A Generalized Information Theory" (in Chinese) [15]. In 1999, he improved this theory and introduced it in an English journal [16]. Since 2017, the author has adopted the P-T probability framework with model parameters for this theory's applications to machine learning [12,13,17].

The semantic information theory introduced in [14,15] is called the semantic information G theory or the G theory for short (G means generalization). The semantic information measure is called the G measure. In the G theory, the information rate-distortion function $R(D)$ is extended to the $R(G)$ function. In $R(G)$, $R$ is the lower limit of ShMI for a given $G$, and $G$ is the upper limit of SeMI for a given $R$. The $R(G)$ function can reveal the relationship between SeMI and ShMI for understanding deep learning. Its applications to machine learning can be found in [12,13,18]. The main difference between semantic information and Shannon's information is that the former involves truth and falsehood, as discussed by Floridi [19]. According to Tarski's semantic conception of truth [20] and Davidson's truth-conditional semantics [21], the truth function of a hypothesis or label determines its semantic meaning (See Appendix A for details).

To measure the semantic information conveyed by a label $y_j$ (a constant) about an instance $x$ (a variable), we need the (fuzzy) truth function (where $x$ makes $y_j$ true). Generally, $x$ and $y_j$ belong to different domains. If $x$ and $y_j$ belong to the same domain, $y_j$ becomes an estimation, i.e., $y_j = \hat{x}_j = $ "$x$ is about $x_j$". In this case, semantic information becomes estimated information. For example, a GPS pointer means an estimate $y_j = \hat{x}_j$ (see Section 3.1 for details); it conveys estimated information. Since it may be wrong, the information is also semantic information. Therefore, we can say that estimated information is a special case of semantic information; this paper also regards estimated information as semantic information and the learning method of using EMI as the semantic ITL method.

Although the method of using estimated information in deep learning is very successful, there are still many problems. Firstly, some essential functions and entropy or information measures have no unified names, such as function $m(x, y) = P(x, y)/[P(x) P(y)]$ and generalized entropies related to EMI. It is particularly worth mentioning that EMI, as the protagonist, does not have its own name. It is often called the lower bound of mutual information. On the other hand, the same name may represent very different functions. For example, the semantic similarity [22] may be expressed with $\exp[-d(x, y)]$ between 0 and 1 or $-\log P(c)$ [23] between 0 and $\infty$. To exchange ideas conveniently, we should unify the names of various functions and entropy and information measures.

Secondly, there are some essential questions that the researchers of deep learning cannot answer. The questions include:

- How are EMI and ShMI related in supervised, semi-supervised, and unsupervised learning?
- Is similarity probability? If it is, why is it not normalized? If it is not, why can we bring it into Bayes' formula (see Equation (9))?
- Can we get similarity functions, distortion functions, truth functions, or membership functions directly from samples or sampling distributions?

However, the G theory can answer these questions clearly.

Thirdly, due to the unclear understanding of the relationship between EMI and ShMI, many deep learning researchers [4,24] believe that maximizing ShMI is a good objective; since EMI is the lower bound of ShMI, we can maximize ShMI by maximizing EMI. However, Tschannen et al. [25] think that maximizing ShMI is not enough or not good in some cases; there are other reasons for the recent success of deep learning. The author thinks that their questioning deserves attention. From the G theory perspective, EMI is similar to the utility, and ShMI is like the cost; our purpose is to maximize EMI. In some cases, both need maximization; while in other cases, to improve communication efficiency, we need to reduce ShMI.

Tishby et al. [26,27] use the information bottleneck to interpret deep learning. According to their theory, for a neural network with the structure $X \rightarrow T \rightarrow Y$, we need to maximize $I(Y; T) - \beta I(X; T)$ [26] or minimize $I(X; T) - \beta I(Y; T)$, where $\beta$ is a Lagrange multiplier that controls the trade-off between maximizing $I(Y; T)$ and minimizing $I(X; T)$. This interpretation involves the rate-distortion function $R(D)$ and is very inspiring. However, in this theory, EMI (or parameterized mutual information) and ShMI are not well distinguished, ignoring that making them close to each other is an important task. Therefore, this paper (Section 5.2) will provide a competitive interpretation using the $R(G)$ function.

Despite various problems in deep learning, the Deep Neural Network (DNN) has a strong feature-extraction ability; combined with the EMI measure, it provides many convincing applications. In contrast, applying the G theory to machine learning is relatively simple without combining DNNs. From the practical viewpoint, the G theory's applications lack persuasiveness. Information theories (including the G theory) are also insufficient to explain many ingenious methods in deep learning.

Now is the time to merge the two trends and let them learn from each other!

There are many generalized entropies. However, the generalized entropies related to the G measure keep the structure of the Shannon entropy. Only the function on the right of the log is changed. In addition to the cross-entropy method for classical ITL [28,29], researchers have also proposed other generalized entropies, such as Reny Entropy [30,31], Tsallis Entropy [32,33], Correntropy [34], matrix-based Entropy [35], and related methods for generalized ITL [35]. Figure 1 shows the distinctions and relations between classical, generalized, and semantic ITL.

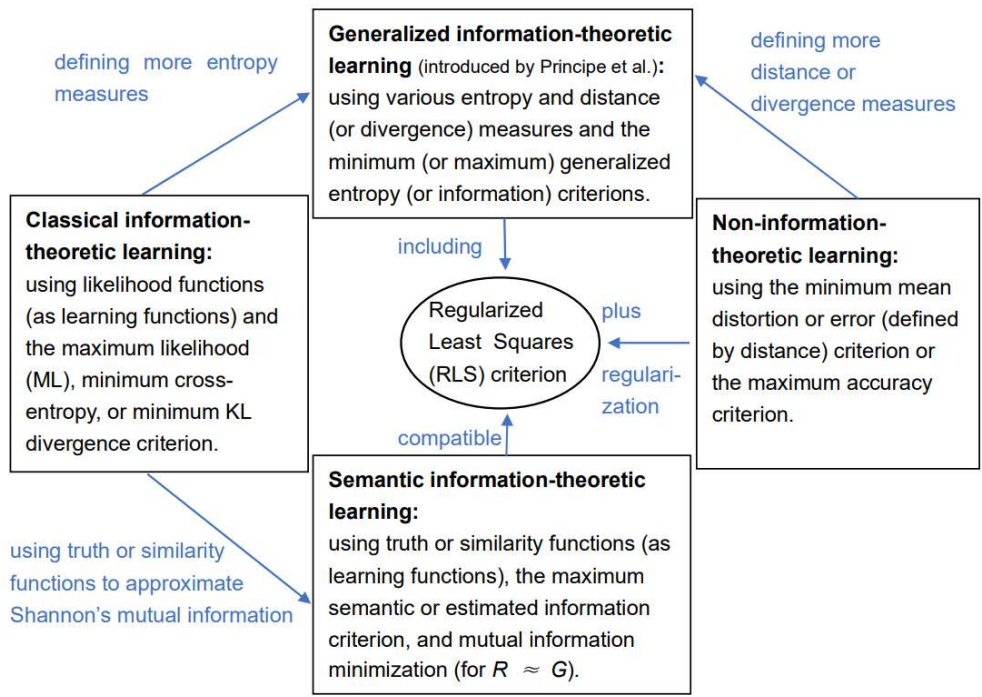

**Figure 1.** The distinctions and relations between four types of learning [31,34].

The main difference between Semantic ITL and Non-ITL is that the latter uses truth, correctness, or accuracy as the evaluation criterion, whereas the former uses truthlikeness (i.e., the successful approximation to truth) [36] or its logarithm as the evaluation criterion. Truth is independent of the prior probability distribution $P(x)$ of instance $x$, whereas truthlikeness is related to $P(x)$. Two criteria should have different uses. The main difference between Semantic ITL and generalized ITL in Figure 1 is that the former only generalizes Shannon's entropy and information with truth functions.

Although many methods use parameters or SoftMax functions to construct mutual information, they cannot be regarded as semantic ITL methods if they do not use similarity or truth functions. Of course, these entropy and information measures also have their application values, but they are not what this article is about.

In addition to the semantic information measures mentioned above, there are other semantic information measures, such as those proposed by Floridi [37] and Zhong [38]. However, the author has not found their applications to machine learning.

This paper mainly aims at:

1. Reviewing the evolutionary histories of semantic information measures and learning functions;
2. Clarifying the relationship between SeMI and ShMI;
3. Promoting the integration and development of the G theory and deep learning.

The contents of the following sections include:

- Reviewing the evolution of semantic information measures;
- Reviewing the evolution of learning functions;
- Introducing the G theory and its applications to machine learning;
- Discussing some questions related to SeMI and ShMI maximization and deep learning;
- Discussing some new methods worth exploring and the limitation of the G theory;
- Conclusions with opportunities and challenges.

**2. The Evolution of Semantic Information Measures**

*2.1. Using the Truth or Similarity Function to Approximate to Shannon's Mutual Information*

The ShMI formula is:

$$I(X;Y) = \sum_i \sum_j P(x_i, y_j) \log \frac{P(x_i, y_j)}{P(x_i)P(y_j)} \quad (1)$$
$$= H(Y) - H(Y|X) = H(X) - H(X|Y).$$

where $x_i$ is an instance, $y_j$ is a label, and $X$ and $Y$ are two random variables.

To illustrate EMI, let us look at the core part of ShMI:

$$I(x; y_j) = \log \frac{P(x, y_j)}{P(x)P(y_j)} \quad (2)$$

where $x$ is a variable, and $y_j$ is a constant. Deep learning researchers [1, 2] found that we could use parameters to construct similarity functions and then use similarity functions to construct EMI that approximates to ShMI. The main characteristic of similarity functions is:

$$S(x; y_j) \propto m(x, y_j) = \frac{P(x, y_j)}{P(x)P(y_j)} \propto P(y_j|x). \quad (3)$$

The above formula also defines $m(x, y_j)$, an important function. However, the author has never seen its name. We call $m(x, y_j)$ the relatedness function now. We can also construct the distortion function $d(x, y_j)$ with parameters and use $d(x, y_j)$ to express the similarity function so that

$$S(x; y_j) = \exp[-d(x, y_j)] \propto m(x, y_j). \quad (4)$$

Then we can construct estimated information:

$$I_\theta(xk; yk) = \log \frac{S(xk, yk)}{\sum_{l=1}^{N} S(xl, yl)} = \log \frac{\exp[-d(xk, yk)]}{\sum_{l=1}^{N} \exp[-d(xl, yl)]}, \quad (5)$$

where $k$ denotes the sequence number of the $k$-th example in a sample $\{(xk, yk) \mid k = 1, 2, ..., N\}$. EMI is expressed as:

$$I_\theta(X;Y) = \frac{1}{N} \sum_{k=1}^{N} I_\theta(xk; yk). \quad (6)$$

On the other hand, information-theoretic researchers generally use sampling distribution $P(x, y)$ to average information. Estimated information and EMI are expressed as

$$I_\theta(x_i; y_j) = \log \frac{S(x_i, y_j)}{\sum_l P(x_l)S(x_l, y_j)} = \log \frac{\exp[-d(x_i, y_j)]}{\sum_l P(x_l)\exp[-d(x_l, y_j)]}, \quad (7)$$

$$I_\theta(X;Y) = \sum_j \sum_i P(x_i, y_j) I_\theta(x_i; y_j). \quad (8)$$

Note that $P(x_l)$ is added in the partition function. Nevertheless, Equations (4) and (6) are equivalent. Compared with the likelihood function $P(x|\theta_j)$ (where $\theta_j$ represents $y_j$ and related parameters), the similarity function is independent of the source $P(x)$. After the source and the channel change, the similarity function is still proper as a predictive model. We can use the similarity function and the new source $P'(x)$ to make a new probability prediction or produce a likelihood function:

$$P'(x|\theta_j) = \frac{P'(x)S(x,y_j)}{\sum_l P'(x_l)S(x_l,y_j)}. \quad (9)$$

Equations (5) and (6) fit small samples, whereas Equations (7) and (8) fit large samples. In addition, the latter can indicate the change in the source and helps clarify the relationship between SeMI and ShMI.

*2.2. Historial Events Related to Semantic Information Measures*

The author lists major historical events related to semantic information measures to the best of his knowledge.

In 1931, Popper put forward in the book "*The Logic of Scientific Discovery*" [39] (p. 96) that the smaller the logical probability of a scientific hypothesis, the greater the amount of (semantic) information if it can stand the test. We can say that Popper is the earliest researcher of semantic information [19]. Later, he proposed a logical probability axiom system. He emphasized that there are two kinds of probabilities, statistical and logical probabilities, at the same time [39] (pp. 252–258). However, he had not established a probability system that includes both.

In 1948, Shannon [3] published his famous paper: *A mathematical theory of communication*. Shannon's information theory provides a powerful guiding ideology for optimizing communication codes. However, Shannon only uses statistical probability based on the frequency interpretation. In the above paper, he also proposed the information rate-fidelity function. It was later renamed as the information rate-distortion function, i.e., $R(D)$ ($D$ is the upper limit of average distortion, and $R$ is the Minimum Mutual Information (MinMI)).

In 1949, Shannon and Weaver [9] jointly published a book: "*The Mathematical Theory of Communication*", which contains Shannon's famous article and Weaver's article: "*Recent Contributions to The Mathematical Theory of Communication*". In Weaver's article, he put forward the three levels of communication:

"Level A. How accurately can the symbols of communication be transmitted?"

"Level B. How precisely do the transmitted symbols convey the desired meaning?"

"Level C. How effectively does the received meaning affect conduct in the desired way?"

Level A only involves Shannon information; Level B is related to both Shannon and semantic information; Level C is also associated with information values.

In the last century, many people rejected the study of semantic information. Their main reason was that Shannon said [9] (p. 3):

"Frequently, the messages have meaning; that is they refer to or are correlated according to some system with certain physical or conceptual entities. These semantic aspect of communication are irrelevant to the engineering problem."

Shannon only limited the application scope of his theory without opposing the study of semantic information. How could he agree with Weaver to jointly publish the book if he objected? In the last ten years, with the developments of artificial intelligence and the Internet, the requirement for semantic communication has become increasingly urgent, and more and more scholars have begun to study semantic communication [10,11], including semantic information theory.

In 1951, Kullback and Leibler [40] proposed Kullback–Leibler (KL) divergence, also known as KL information or relative entropy:

$$D_{KL} = H(P\|Q) = \sum_i P(x_i) \log \frac{P(x_i)}{Q(x_i)}. \quad (10)$$

KL information can be regarded as a special case of ShMI as $Y = y_j$, whereas ShMI can also be seen as the KL information between two distributions: $P(x, y)$ and $P(x)P(y)$.

Since Shannon's information measure cannot measure semantic information, in 1952, Carnap and Bar-Hillel [41] proposed a semantic information formula:

$$I_{CB} = \log[1/m_p]. \quad (11)$$

where $m_p$ is a proposition's logical probability. This formula partly reflects Popper's idea that the smaller the logical probability, the greater the amount of semantic information. However, $I_{CB}$ does not indicate whether the hypothesis can stand the test. Therefore, this formula is not practical. In addition, the logical probability $m_p$ is independent of the prior probability distribution of the instance, which is also unreasonable.

In 1957, Shepard [42] proposed using distance to express similarity: $S(x, y) = \exp[-d(x, y)]$.

In 1959, Shannon [43] provided the solution for a binary memoryless source's $R(D)$ function. He also deduced the parametric solution of $R(D)$ of a general memoryless source:

$$D(s) = \sum_i \sum_j d_{ij} P(x_i) P(y_j) \exp(sd_{ij}) / Z_i,$$
$$R(s) = sD(s) - \sum_i P(x_i) \ln Z_i, \quad (12)$$

where $d_{ij} = d(x_i, y_j)$, $s \leq 0$, and $Z_i$ is the partition function. In the author's opinion, $\exp(sd_{ij})$ is a truth function, and this MinMI can be regarded as SeMI [18].

In 1967, Theil [44] put forward the generalized KL information formula:

$$I_{\text{Theil}} = \sum_i P(x_i) \log \frac{R(x_i)}{Q(x_i)}, \quad (13)$$

where there are three functions. We may regard $R(x)$ as the prediction of $P(x)$ and $Q(x)$ as the prior distribution. Hence, $I_{\text{Theil}}$ means predictive information.

In 1965, Zadeh [45] initiated the fuzzy sets theory. The membership grade of an instance $x_i$ in a fuzzy set $A_j$ is denoted as $M_{Aj}(x_i)$. The author of this paper explained in 1993 [15] that the membership function $M_{Aj}(x)$ is also the truth function of the proposition function $y_j = y_j(x) = $ "$x$ is in $A_j$." If we assume that there is a typical $x_j$ (that is, Plato's idea, which may not be in $A_j$) that makes $y_j$ true, i.e., $M_{Aj}(x_j) = 1$, then the membership function $M_{Aj}(x)$ is the similarity function between $x$ and $x_j$.

In 1972, De Luca and Termini [46] defined fuzzy entropy with a membership function. We call it DT fuzzy entropy.

In 1974, Akaike [47] brought model parameters into the KL information formula and proved that the maximum likelihood estimation is equivalent to the minimum KL information estimation. Although he did not explicitly use the term "cross-entropy", the log-likelihood he uses is equal to the negative cross-entropy multiplied by the sample size $N$:

$$\log P(\mathbf{X}|\theta) = N \sum_i P(x_i) \log P(x_i|\theta) = -NH(X|\theta). \quad (14)$$

In 1981, Thomas [48] used an example to explain that we may bring a membership function into Bayes' formula to make a probability prediction (see Equation (31)). According to Dubois and Prade's paper [49], other people almost simultaneously mentioned such a formula.

In 1983, Donsker and Varadhan [50] put forward that KL information can be expressed as:

$$I_{DV} = \sum_i P(x_i) T(x_i) - \log \sum_k Q(x_k) \exp[T(x_k)] \leq H(P\|Q) = D_{KL} \quad (15)$$

$I_{DV}$ was later called the Donsker–Varadhan representation in [1]. We also called it DV-KL information. They proposed this formula perhaps because they were inspired by the in-formation rate-distortion function or the Gibbs (or Boltzmann–Gibbs) distribution. To understand this formula, we replace $P(x)$ with $P(x|y_j)$ and $Q(x)$ with $P(x)$. Then the KL information becomes:

$$I(X; y_j) = \sum_i P(x_i | y_j) \log \frac{P(x_i | y_j)}{P(x_i)} = \sum_i P(x_i | y_j) \log \frac{P(y_j | x_i)}{P(y_j)}. \qquad (16)$$

To express KL information, we only need to find $T(x_i)$ so that $\exp[T(x_i)] \propto P(y_j|x)$. DV-KL information was later used for MINE [1]. However, exponential or negative-exponential functions are generally symmetrical, while $P(y_j|x)$ is usually asymmetrical. Thus, it is not easy to make two information quantities equal.

In 1983, Wang [51] proposed the statistical interpretation of the membership function by the Random Set Falling Shadow theory. In terms of machine learning, we can use a sample in which each example contains a label and some similar instances, that is, $S = \{(x_{t1}, x_{t2}, …; y_t) | t = 1, 2, …, N)\}$, the probability of $x_i$ in examples with label $y_j$ is membership grade $M_{Aj}(x_i)$. The author later [17] proved that the membership function $M_{Aj}(x)$ is a truth function and can be obtained from a regular sample, where an example only includes one instance.

In 1986, Aczel and Forte [52] proposed a generalized entropy, which is now called cross-entropy. They proved an inequality:

$$-\sum_i P_i \log Q_i \geq -\sum_i P_i \log P_i = H(P), \qquad (17)$$

with $\ln x \leq x - 1$. They explain that this formula also reflects Shannon's lossless coding theorem, which means the Shannon entropy is the lower limit of the average codeword length.

In 1986, Zadeh defined the probability of fuzzy events [53], which is equal to the average of a membership function. This paper will show that this probability is the logical probability and the partition function we often use.

In 1990, the author [14] proposed using both the truth function and the logical probability to express semantic information (quantity). This measure can indicate whether a hypothesis can stand the test and overcome the shortcoming of Carnap–Bar-Hillel's semantic information measure. The author later called this measure the semantic information G measure to distinguish it from other semantic information measures. The formula is:

$$I_{LU}(x; y_j) = \log \frac{Q(A_j | x)}{Q(A_j)}, \qquad (18)$$

where $A_j$ is a fuzzy set, $y_j$ = "$x$ belongs to $A_j$" is a hypothesis, and $Q(A_j|x)$ is the membership function of $A_j$ and the (fuzzy) truth function of $y_j$. $Q(A_j)$ is the logical probability of $y_j$ and the probability of fuzzy events called by Zadeh [53]. If there is always $Q(A_j|x) = 1$, the above semantic information formula becomes Carnap–Bar-Hillel's semantic information formula.

In fact, the author first wanted to measure the information of color perceptions. He proposed the decoding model of color vision before [54,55]. This model has three pairs of opponent colors: red-green, blue-yellow, and green-magenta, whereas, in the famous zone model of color vision, there are only two pairs: red-green and yellow-blue. To defend the decoding model by showing that higher discrimination can convey more information, he tried measuring the information of color perceptions. For this reason, he regarded a color perception as the estimate of a color, i.e., $y_j = \hat{x}_j$, and employed the Gaussian function as the discrimination or similarity function. Later, he found that we could also measure natural language information by replacing the similarity function with the truth function. Since statistical probability is used for averaging, this infor-mation measure can ensure that wrong hypotheses or estimates will reduce semantic information.

Averaging $I_{Lu}(x; y_j)$, we obtain generalized Kullback–Leibler (KL) information (if we use $P(x|A_j)$), as proposed by Theil, and semantic KL information (if we use $Q(A_j|x)$), namely:

$$I_{LU}(X;y_j) = \sum_i P(x_i | y_j)\log \frac{P(x_i | A_j)}{P(x_i)} = \sum_i P(x_i | y_j)\log \frac{Q(A_j | x_i)}{Q(A_j)}. \quad (19)$$

The above Equation adopts Bayes' formula with the membership function (see Section 4.1 for details). At that time, the author did not know of Thomas and others' studies. If $Q(A_j|x)$ is expressed as $\exp[-d(x, y_j)]$, semantic KL information becomes DV-KL information:

$$I_{LU}(X;y_j) = -\sum_i P(x_i | y_j)d(x_i, y_j) + \log[1/Q(A_j)]. \quad (20)$$

Nevertheless, the author only later learned of Theil's formula and recently learned of DM-KL information. However, not all KL information or semantic KL information can be expressed as DV-KL information with exponential or negative exponential functions. For example, the truth function of $y_j$ = "Elderly" is asymmetric and can be described by a Logistic function rather than an exponential function. Therefore, DV-KL information is only a particular case of semantic KL information.

In 1993, the author [15] extended Shannon's information rate-distortion function $R(D)$ by replacing $d(x, y)$ with $I_{Lu}(x; y)$, to obtain the information rate-fidelity function $R(G)$, which means minimum ShMI for given SeMI. The $R(G)$ function reveals the matching relationship between ShMI and SeMI. Meanwhile, he defined the truth value matrix as the generalized channel and proposed the idea of mutual matching of two channels. He also studied how to compress image data according to the visual discrimination of colors and the $R(G)$ function [15,16].

In 2017, the author proposed the P-T probability framework [12] by replacing $Q$ with $T$ and $A_j$ with $\theta_j$. In 2020, He discussed how the P-T probability framework is related to Popper's theory [17]. The $\theta$ represents not only a fuzzy set but also a set of model parameters. With the P-T probability framework, we can conveniently use the G measure for machine learning. The relationship between SeMI and several generalized entropies is:

$$I(X;Y_\theta) = \sum_i \sum_j P(x_i)P(y_j | x_i)\log \frac{T(\theta_j | x_i)}{T(\theta_j)} \quad (21)$$
$$= H(Y_\theta) - H(Y_\theta | X) = H(X) = H(X | Y_\theta),$$

where three generalized entropies are:

$$H(X|Y_\theta) = -\sum_j \sum_i P(x_i, y_j)\log P(x_i | \theta_j), \quad (22)$$

$$H(Y_\theta | X) = -\sum_j \sum_i P(x_i, y_j)\log T(\theta_j | x_i), \quad (23)$$

$$H(Y_\theta) = -\sum_j P(y_j)\log T(\theta_j). \quad (24)$$

$H(X|Y_\theta)$ can be called the prediction entropy. It is also a cross-entropy. About the other two generalized entropies, the author suggests that we call $H(Y_\theta|X)$ the fuzzy entropy and $H(Y_\theta)$ the coverage entropy (see Section 5.4 for details).

Compared with the DT fuzzy entropy, $H(Y_\theta|X)$ as the fuzzy entropy is more general. The reason is that if $n$ labels become two complementary labels, $y_1$ and $y_0$, and $P(y_1|x)$ and $P(y_0|x)$ equals two membership functions $T(\theta_1|x)$ and $T(\theta_0|x)$, fuzzy entropy $H(Y_\theta|X)$ degenerates into the DT fuzzy entropy.

More fuzzy entropies and fuzzy information measures defined by others can be found in [56]. The author thinks that some semantic information measures constructed with the DT fuzzy entropy are inappropriate (see the analysis of Equation (1) in [13]). The main reason is that they cannot reflect whether a hypothesis can stand the test.

In 2021, the author extended the information rate-distortion function and the maximum entropy method by replacing distortion function $d(x, y_j)$ with $-\log T(\theta_j|x)$ and the average distortion with $H(Y_\theta|X)$ [18]. In this way, the constraint is more flexible.

In recent years, deep learning researchers have expressed similarity functions with feature vectors and brought similarity functions into semantic or estimated information measures. Their efforts have made semantic information measures more abundant and more practical. See the next section for details.

## 3. The Evolution of Learning Functions

### 3.1. From Likelihood Functions to Similarity and Truth Functions

This section reviews machine learning with different learning functions and criteria for readers to understand why we choose the similarity (or truth) function and the estimated (or semantic) information measure.

The task of machine learning is to use samples or sampling distributions to optimize learning functions and then use learning functions to make probability predictions or classifications. There are many criteria for optimizing learning functions, such as maximum accuracy, maximum likelihood, Maximum Mutual Information (MaxMI), maximum EMI, minimum mean distortion, Regularized Least Squares (RLS), minimum KL divergence, and minimum cross-entropy criteria. Some criteria are equivalent, such as minimum cross-entropy and maximum EMI criteria. In addition, some criteria are similar or compatible, such as the RLS and the maximum EMI criteria.

Different Learning functions approximate to the following different probability distributions:

- $P(x)$ is the prior distribution of instance $x$, representing the source. We use $P_\theta(x)$ to approximate to it.
- $P(y)$ is the prior distribution of label $y$, representing the destination. We use $P_\theta(y_j)$ to approximate to it.
- $P(x|y_j)$ is the posterior distribution of $x$. We use the likelihood function $P(x|\theta_j) = P(x|y_j, \theta)$ to approximate to it.
- $P(y_j|x)$ is the transition probability function [13]. What approximates to it is $P(\theta_j|x)$. Fisher proposed it and called it the inverse probability [57,58]. We call it the inverse probability function. The Logistic function is often used in this way.
- $P(y|x_i)$ is the posterior distribution of $y$. Since $P(y|x_i) = P(y)P(x|y)/P(x_i)$, Bayesian Inference uses $P(\theta) P(x|y, \theta)/P_\theta(x)$ (Bayesian posterior) [57] to approximate to it.
- $P(x, y)$ is the joint probability distribution. We use $P(x, y|\theta)$ to approximate to it.
- $m(x, y_j) = P(x, y_j)/[P(x)P(y_j)]$ is the relatedness function. What approximates to it is $m_\theta(x, y_j)$. We call $m_\theta(x, y_j)$ the truthlikeness function, which changes between 0 and $\infty$.
- $m(x, y_j)/\max[m(x, y_j)] = P(y_j|x)/\max[P(y_j, x)]$ is the relative relatedness function. We use the truth function $T(\theta_j|x)$ or the similarity function $S(x, y_j)$ to approximate to it.

The author thinks that learning functions have evolved from likelihood functions to truthlikeness, truth, and similarity functions. Meanwhile, the optimization criterion has also evolved from the maximum likelihood criterion to the maximum semantic (or estimated) information criterion. In the following, we investigate the reasons for the evolution of learning functions.

We cannot use a sampling distribution $P(x|y_j)$ to make a probability prediction because it may be unsmooth or even intermittent. For this reason, Fisher [57] proposed using the smooth likelihood function $P(x|\theta_j)$ with parameters to approximate to $P(x|y_j)$ with the maximum likelihood criterion. Then we can use $P(x|\theta_j)$ to make a probability prediction.

The main shortcoming of the maximum likelihood method is that we have to retrain $P(x|\theta_j)$ after $P(x)$ is changed. In addition, $P(x|y_j)$ is often irregular and difficult to be approximated by a function. Therefore, the inverse probability function $P(\theta_j|x)$ is used. With this function, when $P(x)$ becomes $P'(x)$, we can obtain a new likelihood function $P'(x|\theta_j)$ by using Bayes' formula. In addition, we can also use $P(\theta_j|x)$ for classification with the maximum accuracy criterion.

The Logistic function is often used as the inverse probability function. For example, when $y$ has only two possible values, $y_1$ and $y_0$, we can use a pair of Logistic functions with parameters to approximate to $P(y_1|x)$ and $P(y_0|x)$. However, this method also has two disadvantages:

- When the number of labels is $n > 2$, it is difficult to construct a set of inverse probability functions because $P(\theta_j|x)$ should be normalized for every $x_i$:

$$\sum_j P(\theta_j | x_i) = 1, \text{ for } i = 1, 2, ... \tag{25}$$

  The above formula's restriction makes applying the inverse probability function to multi-label learning difficult. An expedient method is Binary Relevance [59], which collects positive and negative examples for each label, and converts a multi-label learning task into $n$ single-label learning tasks. However, preparing $n$ samples for $n$ labels is uneconomical, especially for image and text learning. In addition, the results are different whether a group of inverse probability functions are optimized separately or together.

- Although $P(y_j|x_i; i = j)$ indicates the accuracy for binary communication, when $n > 2$, $P(y_j|x_i; i = j)$ may not mean it, especially for semantic communication. For example, $x$ represents an age, $y$ denotes one of the three labels: $y_0$ = "Non-adult", $y_1$ = "Adult", and $y_2$ = "Youth". If $y_2$ is rarely used, both $P(y_2)$ and $P(y_2|x)$ are tiny. However, the accuracy of using $y_2$ for $x$ = 20 should be 1.

$P(x, y|\theta)$ is also often used as a learning function, such as in the RBM [60] and many information-theoretic learning tasks [29]. Like $P(x|\theta_j)$, $P(x, y|\theta)$ is not suitable for irregular or changeable $P(x)$. The Bayesian posterior is also unsuitable.

Therefore, we need a learning function that is proportional to $P(y_j|x)$ and independent of $P(x)$ and $P(y)$. The truth function and the similarity function are such functions.

We use GPS as an example to illustrate why we use the truth (or similarity) function $T(\theta_j|x)$, instead of $P(x|\theta_j)$ or $P(x, y|\theta)$, as the learning function. A GPS pointer indicates an estimate $y_j = \hat{x}_j$, whereas the real position of $x$ may be a little different. In this case, the similarity between $x$ and $\hat{x}_j$ is the truth value of $\hat{x}_j$ as $x$ happens. This explains why the similarity is often called the semantic similarity [22].

Figure 2 shows a real screenshot, where the dashed line, the red star, and letters are added. The brown line represents the railway; the GPS device is in a train on the railway. Therefore, we may assume that $P(x)$ is equally distributed over the railway. The $y_j$ does not point at the railway because it is slightly inaccurate.

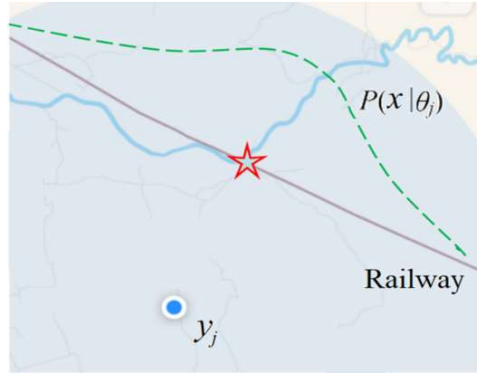

**Figure 2.** Illustrating a GPS device's positioning with a deviation. We predict the probability distribution of $x$ according to $y_j$ and the prior knowledge $P(x)$. The red star represents the most probable position.

The semantic meaning of the GPS pointer can be expressed with

$$T(\theta_j|x) = S(x, x_j) = \exp[-(x - x_j)^2/(2\sigma^2)], \tag{26}$$

where $x_j$ is the pointed position by $y_j$, and $\sigma$ is the Root Mean Square (RMS). For simplicity, we assume $x$ is one-dimensional in the above Equation. According to Equation (9), we can predict that the star represents the most probable position.

However, if we use $P(x|\theta_j)$ or $P(x, y|\theta_j)$ as the learning function and the predictive model, after $P(x)$ is changed, this function cannot make use of the new $P(x)$, and hence the learning is not transferable. In addition, using the truth function, we can learn the system deviation and the precision of a GPS device from a sample [13]. However, using $P(x|\theta_j)$ or $P(x, y|\theta_j)$, we cannot do this because the deviation and precision are independent of $P(x)$ and $P(y)$. Consider a car with a GPS map on a road, where $P(x)$ and $P(x|\theta_j)$ may be more complex and changeable. However, $T(\theta_j|x)$ can be an invariant Gaussian function.

The above example explains why we use similarity functions as learning functions. Section 4.2 proves that the similarity function approximates to $m(x, y_j)/\max[m(x, y_j)]$.

### 3.2. The Definitions of Semantic Similarity

"Similarity" in this paper means semantic similarity [22]. For example, people of ages 17 and 18, snow and water, or message and information are similar. However, it does not include superficial similarities, such as between snow and salt or between "message" and "massage".

Semantic similarity can be between two sensory, geometric, biological, linguistic, etc., objects. The similarity is often defined by distance. Distance also reflects distortion or error. Following Shepard [42], Ashby and Perrin [61] further discussed the probability prediction related to similarity, distance, and perception. The similarity between images or biological features is usually expressed with the dot product or cosine of two feature vectors [2,4,6].

There are many studies on the similarity between words, sentences, and texts [22]. Some researchers use semantic structure to define the similarity between two words. For example, in 1995, Resnik [23] used the structure of WordNet to define the similarity between any two words. Later, others proposed some improved forms [62]. Some researchers use statistical data to define similarity. For example, in Latent Semantic Analysis [63], researchers use vectors from statistics to express similarity, whereas, some researchers use Pointwise Mutual Information (PMI) [64], i.e., $I(x_i; y_j)$ (see Equation (2)), as a similarity measure. Since PMI varies between $-\infty$ and $\infty$, an improved PMI similarity between 0 and 1 was proposed [65]. We can also see that semantic similarity is defined with semantic distance. More semantic similarity and semantic distance measures can be found in [66]. Some researchers believe semantic similarity includes relatedness, while others believe the two are different [67]. For example, hamburgers and hot dogs are similar, while hamburgers and French fries are related. However, the author believes that because the relatedness of two objects reflects the similarity between their positions in space or time, it is practical and reasonable to think that the related two objects are similar.

The truth function $T(\theta_j|x)$ proposed by the author can also be understood as the similarity function (between $x$ and $x_j$). We can define it with distance or distortion and optimize it with a sampling distribution. The author uses $T(\theta_j|x)$ to approximate to $m(x, y_j)/\max[m(x, y_j)]$. When the sample is large enough, they are equal.

The similarity function $S(x, y)$ can be directly used as a learning function to generate the posterior distribution of $x$. We can use such a transformation: $\exp[kS(x, y)]$, as a new similarity function or truth function [2,6] (where $k$ is a parameter that can enlarge or reduce the similarity range). For those similarity measures suitable to be put in $\exp(\ )$, such as the dot product or the cosine of two vectors, we had better understand them as fidelity measures and denote them as $f_d(x_i, y_j)$. Assuming that the maximum possible fidelity is $f_{d\max}$, we may define the distortion function: $d(x, y_j) = k[f_{d\max} - f_d(x, y_j)]$, and then use $\exp[-d(x, y_j)]$ as the similarity function. Nevertheless, with or without this conversion, the value of the SoftMax function or the truthlikeness function is unchanged.

The study of semantic similarity reminds us that we should use similarity functions to measure semantic information (including sensory information). In machine learning, it is important to use parameters to construct similarity functions and then use samples or sampling distributions to optimize the parameters.

*3.3. Similarity Functions and Semantic Information Measures Used for Deep Learning*

In 1985, Achley, Hinton, and Sejnowski proposed the Boltzmann machine [68], and later they proposed the Restricted Boltzmann Machine (RBM) [60]. The SoftMax function is used to generate the Gibbs distribution. In 2006, Hinton et al. used the RBM and Back-propagation to optimize the AutoEncoder [69] and the Deep Belief Net (DBN) [70], achieving great success. However, the RBM is a classical ITL method, not a semantic ITL method, because the SoftMax function approximates to $P(x, y)$ rather than a truthlikeness function with a similarity function. Nevertheless, using the SoftMax function demonstrates the use of similarity functions later.

In 2010, Gutmann and Hyvärinen [71] proposed noise contrast learning and obtained good results. They used non-normalized functions and partition functions. From the author's viewpoint, if we want to get $P(\theta_j|x)$ for a given $P(x|y_j)$, we need $P(x)$ because $P(y_j|x)/P(y_j) = P(x|y_j)/P(x)$. Without $P(x)$, we can create a counterexample by noise and get $P(x)$. Then we can optimize $P(\theta_j|x)$. Like Binary Relevance, this method converts a multi-label learning task into $n$ single-label learning tasks.

In 2016, Sohn [72] creatively proposed distance metric learning with the SoftMax function. Although it is not explicitly stated that the learning function is proportional to $m(x, y)$, the distance is independent of $P(x)$ and $P(y)$.

In 2017, the author concluded [12] that when $T(\theta_j|x) \propto P(x|y_j)/P(x) = m(x, y)$ or $T(\theta_j|x) \propto P(y_j|x)$, SeMI reaches its maximum and is equal to ShMI. $T(\theta_j|x)$ is the longitudinal normalization of parameterized $m(x, y)$. When the sample is large enough, we can obtain the optimized truth function from the sampling distribution:

$$T(\theta_j|x) = \frac{m_\theta(x, y_j)}{\max[m_\theta(x, y_j)]} = \frac{m(x, y_j)}{\max[m(x, y_j)]}$$
$$= \frac{P(x, y_j)}{P(x)P(y_j)} \bigg/ \max\left(\frac{P(x, y_j)}{P(x)P(y_j)}\right) = \frac{P(y_j|x)}{\max[P(y_j|x)]}. \quad (27)$$

In addition, the author defined the semantic channel (previously called the generalized channel [15]) with a set of truth functions: $T(\theta_j|x)$, $j$ = 1, 2, ... In [12,13], he developed a group of Channels Matching algorithms for solving the multi-label classification, the MaxMI classification of unseen instances, and mixture models. He also proved the convergence of mixed models [73] and derived the new Bayesian confirmation and causal confirmation measures [74,75]. However, the author has not provided the semantic ITL method's applications to neural networks.

In January 2018, Belghazi et al. [1] proposed MINE, which showed promising results. The EMI for MINE is the generalization of DV-KL information, in which the learning function is:

$$\exp[T_w(x, y_j)] \propto P(y_j | x). \tag{28}$$

Although $T_w(x, y_j)$ is not negative, it can be understood as a fidelity function.

In July 2018, Oord et al. [2] presented InfoNCE and explicitly pointed out that the learning function is proportional to $m(x, y) = P(x|y_j)/P(x)$. The expression in their paper is:

$$f_k(x_{t+k}, c_t) \propto P(x_{t+k} | c_t) / P(x_{t+k}), \tag{29}$$

where $c_t$ is the feature vector obtained from previous data, $x_{t+k}$ is the predictive vector, and $f_k(x_{t+k}, c_t)$ is a similarity function (between predicted $x_{t+k}$ and real $x_{t+k}$) used to construct EMI. The $n$ pairs of Logistic functions in Noise Contrast Learning become $n$ SoftMax functions, which can be directly used for multi-label learning. However, $f_k(x_{t+k}, c_t)$ is not limited to exponential or negative exponential functions, unlike the learning function in MINE. Therefore, the author believes that it is more flexible to use a function such as a membership function as the similarity function.

The estimated information measure for MINE and InfoNCE are almost the same as the author's semantic information G measure. Their characteristics are:

- A function proportional to $m(x, y_j)$ is used as the learning function (denoted as $S(x, y_j)$); its maximum is generally 1, and its average is the partition function $Z_j$.
- The semantic or estimated information between $x$ and $y_j$ is $\log[S(x, y_j)/Z_j]$.
- The statistical probability distribution $P(x, y)$ is used for the average.
- The semantic or estimated mutual information can be expressed as the coverage entropy minus the fuzzy entropy, and the fuzzy entropy is equal to the average distortion.

## 4. The Sematic Information G Theory and Its Applications to Machine Learning

### 4.1. The P-T Probability Framework and the Semantic Information G Measure

In the P-T probability framework, the logical probability is denoted by $T$, and the statistical probability (including subjective probability, such as likelihood) is still represented by $P$. We define this framework as follows.

Let $X$ be a random variable and denote an instance. It takes a value $x \in U = \{x_1, x_2, …\}$. Let $Y$ be a random variable representing a label or hypothesis. It takes a value $y \in V = \{y_1, y_2, ...\}$. A set of transition probability represents a Shannon channel $P(y_j|x)$ ($j$ = 1, 2, ...), whereas a set of truth functions $T(y_j|x)$ ($j$ = 1, 2, …) denotes a semantic channel.

Let elements in $U$ that make $y_j$ true form a fuzzy subset $\theta_j$ (i.e., $y_j$ = "$x$ is in $\theta_j$"). Then the membership function (denoted as $T(\theta_j|x)$) of $x$ in $\theta_j$ is the truth function $T(y_j|x)$ of proposition function $y_j(x)$. That is $T(\theta_j|x) = T(y_j|x)$. The logical probability of $y_j$ is the probability of a fuzzy event defined by Zadeh [45] as:

$$T(y_j) = T(\theta_j) = \sum_i P(x_i)T(\theta_j | x_i). \tag{30}$$

When $y_j$ is true, the predicted probability of $x$ is

$$P(x|\theta_j) = P(x)T(\theta_j|x)/T(\theta_j), \qquad (31)$$

where $\theta_j$ can also be regarded as the model parameter so that $P(x|\theta_j)$ is a likelihood function.

For estimation, $U = V$ and $y_j = \hat{x}_j =$ "$x$ is about $x_j$." Then $T(\theta_j|x)$ can be understood as the confusion probability or similarity between $x$ and $x_j$. For given $P(x)$ and $P(x|\theta_j)$, supposing the maximum of $T(\theta_j|x)$ is 1, we can derive [13]:

$$T(\theta_j|x) = \frac{P(x|\theta_j)T(\theta_j)}{P(x)}, \quad T(\theta_j) = 1/\max[P(x|\theta_j)/P(x)] \qquad (32)$$

The main difference between logical and statistical probabilities is that statistical probability is horizontally normalized (the sum is 1), while logical probability is longitudinally normalized (the maximum is 1). Generally, the maximum of each $T(\theta_j|x)$ is 1 (for different $x$), and $T(\theta_1) + T(\theta_2) + ... > 1$. In contrast, $P(y_0|x) + P(y_1|x) + ... = 1$ for every $x$, and $P(y_1) + P(y_2) + ... = 1$.

Semantic information (quantity) conveyed by (true) $y_j$ about $x_i$ is:

$$I(x_i;\theta_j) = \log\frac{P(x_i|\theta_j)}{P(x_i)} = \log\frac{T(\theta_j|x_i)}{T(\theta_j)} \qquad (33)$$

This Semantic information measure, illustrated in Figure 3, can indicate whether a hypothesis can stand the test.

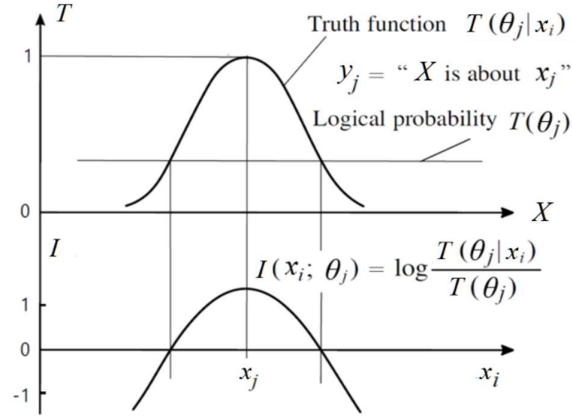

**Figure 3.** The semantic information conveyed by $y_j$ about $x_i$ decreases with the deviation or distortion increasing. The larger the deviation is, the less information there is.

Averaging $I(x; \theta_j)$ for different $x$, we obtain semantic KL information $I(X; \theta_j)$ (see Equation (19)). Averaging $I(X; \theta_j)$ for different $y$, we get SeMI $I(X; Y_\theta)$ (See Equation (21)).

*4.2. Optimizing Truth Functions and Making Probability Predictions*

A set of truth functions, $T(\theta_j|x)$ ($j = 1, 2, ...$), constitutes a semantic channel, just as a set of transition probability functions, $P(y_j|x)$ ($j = 1, 2, ...$), forms a Shannon channel. When the semantic channel matches the Shannon channel, that is, $T(\theta_j|x) \propto P(y_j|x) \propto P(x|y_j)/P(x)$, or $P(x|\theta_j) = P(x|y_j)$, the semantic KL information and SeMI reach their maxima. If the sample is large enough, we have:

$$T(\theta_j|x) = \frac{m(x,y_j)}{mm_j}, \quad m(x,y_j) = \frac{P(x,y_j)}{P(x)P(y_j)}, \quad mm_j = \max[m(x,y_j)] \qquad (34)$$

where $mm_j$ is the maximum of function $m(x, y_j)$ (for different $x$). The author has proved in [17] that the above formula is compatible with Wang's Random Set Falling Shadow theory [51]. We can think that $T(\theta_j|x)$ in Equation (34) comes from Random Point Falling Shadow.

We call $m(x, y_j)$ the relatedness function, which varies between 0 and ∞. Note that relatedness functions are symmetric, whereas truth functions or membership functions are generally asymmetric, i.e., $T(\theta_{xi}|y_j) \neq T(\theta_j|x_i)$ ($\theta_{xi}$ means $x_i$ and related parameters). The reason is that $mm_i = \max[m(x_i, y)]$ is not necessarily equal to $mm_j = \max[m(x, y_j)]$. If we replace $mm_j$ and $mm_i$ with the maximum in matrix $m(x, y)$, the truth function is also symmetrical, like the distortion function. In that case, the maximum of a truth function may be less than 1, so it is not convenient to use a negative exponential function to express a similarity function. Nevertheless, the similarity function $S(x, \hat{x}_j)$ between different instances should be symmetrical, i.e., $S(x_j, \hat{x}_i) = S(x_i, \hat{x}_j)$. The truth function expressed as $\exp[-d(x, y)]$ should also be symmetrical if $d(x, y)$ is symmetrical.

The author thinks that $m(x, y_j)$ or $m(x, y)$ is an essential function. From the perspective of calculation, there exists $P(x, y)$ before $m(x, y)$; but from a philosophical standpoint, there exists $m(x, y)$ before $P(x, y)$. Therefore, we use $m_\theta(x, y)$ to approximate to $m(x, y)$ and call $m_\theta(x, y)$ the truthlikeness function.

If the sample is not large enough, we can use the semantic KL information formula to optimize the truth function:

$$T^*(\theta_j | x) = \underset{T(\theta_j|x)}{\arg\max} \sum_i P(x_i | y_j) \log \frac{T(\theta_j | x_i)}{T(\theta_j)}. \tag{35}$$

Using $m(x, y)$ for probability predictions is simpler than using Bayes' formula because

$$P(x | y_j) = P(x) m(x, y_j), \quad P(y | x_i) = m(x_i, y) P(y). \tag{36}$$

Using $m_\theta(x, y)$ is similar. Unfortunately, it is difficult for the human brain to remember truthlikeness functions. Nevertheless, it is easier for the human brain to remember truth functions. Therefore, we need $T(y_j|x)$ or $S(x, y_j)$. With the truth or similarity function, we can also make probability predictions when $P(x)$ is changed (see Equation (31)).

*4.3. The Information Rate-Fidelity Function R(G)*

To extend the information rate-distortion function $R(D)$, we replace the distortion function $d(x, y_j)$ with semantic information $I(x; \theta_j)$ and replace the upper limit $D$ of average distortion $\bar{d}$ with the lower limit $G$ of SeMI. Then $R(D)$ becomes the information rate-fidelity function $R(G)$ [13,18] (see Figure 4). Finally, following the deduction for $R(D)$, we obtain the $R(G)$ function with parameter $s$:

$$G(s) = \sum_i \sum_j P(x_i) P(y_j | x_i) I_{ij} = \sum_i \sum_j I_{ij} P(x_i) P(y_j) \left[ m_\theta(x_i, y_j) \right]^s / Z_i,$$

$$R(s) = sG(s) - \sum_i P(x_i) \log Z_i, \quad Z_i = \sum_k P(y_k) \left[ m_\theta(x_i, y_j) \right]^s. \tag{37}$$

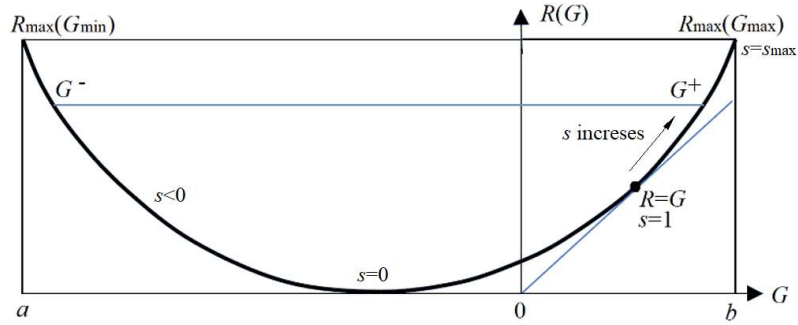

**Figure 4.** The information rate-fidelity function $R(G)$ for binary communication. Any $R(G)$ function is a bowl-like function. There is a point at which $R(G) = G$ ($s = 1$). For given $R$, two anti-functions exist: $G^-(R)$ and $G^+(R)$.

To get the Shannon channel that matches the constraints, we need iterations for the MinMI. That is to repeat the following two formulas:

$$P(y_j | x_i) = P(y_j)\left[m_\theta(x_i, y_j)\right]^s / Z_i, \ i = 1, 2, ...; \ j = 1, 2, ... \\ P(y_j) = \sum_i P(x_i) P(y_j | x_i). \tag{38}$$

In the $R(G)$ function, $s = dR/dG$ is positive on the right side, whereas $s$ in the $R(D)$ function is always negative. When $s = 1$, $R$ equals $G$, meaning that the semantic channel matches the Shannon channel. $G/R$ indicates information efficiency. We can apply the $R(G)$ function to image compression according to visual discrimination [16], semantic compression [18], and the convergence proofs for the MaxMI classification of unseen instances [13] and mixture models [73]. In addition, the author proves that the MaxMI test is equivalent to the maximum likelihood ratio test [76]. More discussions about the $R(G)$ function can be found in [13,18].

*4.4. Channels Matching Algorithms for Machine Learning*

4.4.1. For Multi-Label Learning

We consider multi-label learning, which is supervised learning. We can obtain the sampling distribution $P(x, y)$ from a sample. If the sample is large enough, we let $T(\theta_j|x) = P(y_j|x)/\max[P(y_j|x)]$, $j = 1, 2, ...$; otherwise, we can use the semantic KL information formula to optimize $T(\theta_j|x)$ (see Equation (35)).

For multi-label classifications, we can use the following classifier:

$$y_j^* = h(x) = \arg\max_{y_j} I(x; \theta_j) = \arg\max_{y_j} \log \frac{T(\theta_j | x)}{T(\theta_j)}. \tag{39}$$

Binary Relevance [59] is unnecessary. See [13] for details.

4.4.2. For the MaxMI Classification of Unseen Instances

This classification belongs to semi-supervised learning. Figure 5 illustrates a simple example showing a medical test or signal detection. In Figure 5, $Z$ is a random variable taking a value $z \in C$. The probability distributions $P(x)$ and $P(z|x)$ are given. The task is to find the classifier $y = h(z)$ that maximizes ShMI between $X$ and $Y$.

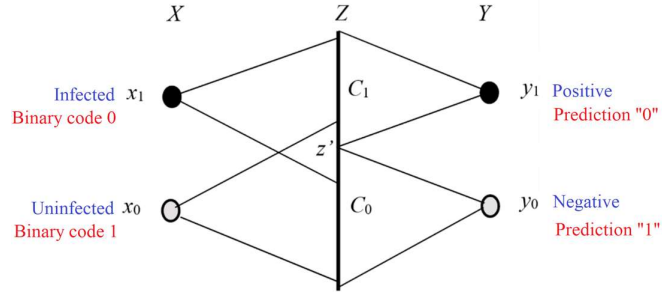

**Figure 5.** Illustrating the medical test and the signal detection. We choose $y_j$ according to $z \in C_j$. The task is to find the dividing point $z'$ that results in MaxMI between $X$ and $Y$.

The following algorithm is not limited to binary classifications. Let $C_j$ be a subset of $C$ and $y_j = f(z | z \in C_j)$; hence $S = \{C_1, C_2, ...\}$ is a partition of $C$. Our task is to find the optimized $S$, which is

$$S^* = \arg\max_S I(X; Y_\theta | S) = \arg\max_S \sum_j \sum_i P(C_j) P(x_i | C_j) \log \frac{T(\theta_j | x_i)}{T(\theta_j)} \quad (40)$$

First, we initiate a partition. Then we do the following iterations.

**Matching I**: Let the semantic channel match the Shannon channel and set the reward function. First, for a given $S$, we obtain the Shannon channel:

$$P(y_j | x) = \sum_{z_k \in C_j} P(z_k | x), j = 1, 2, ..., n. \quad (41)$$

Then we obtain the semantic channel $T(y|x)$ from the Shannon channel and $T(\theta_j)$ (or $m_\theta(x, y) = m(x, y)$). Then we have $I(x_i; \theta_j)$. For given $z$, we have conditional information as the reward function:

$$I(X; \theta_j | z) = \sum_i P(x_i | z) I(x_i; \theta_j), j = 0, 1, ..., n. \quad (42)$$

**Matching II:** Let the Shannon channel match the semantic channel by the classifier:

$$y_j^* = f(z) = \arg\max_{y_j} I(X; \theta_j | z), j = 0, 1, ..., n. \quad (43)$$

Repeat **Matching I** and **Matching II** until $S$ does not change. Then, the convergent $S$ is $S^*$ we seek. The author has explained the convergence with the $R(G)$ function (see Section 3.3 in [13]).

Figure 6 shows an example. The detailed data can be found in Section 4.2 of [13]. The two lines in Figure 6a represent the initial partition. Figure 6d shows that the convergence is very fast.

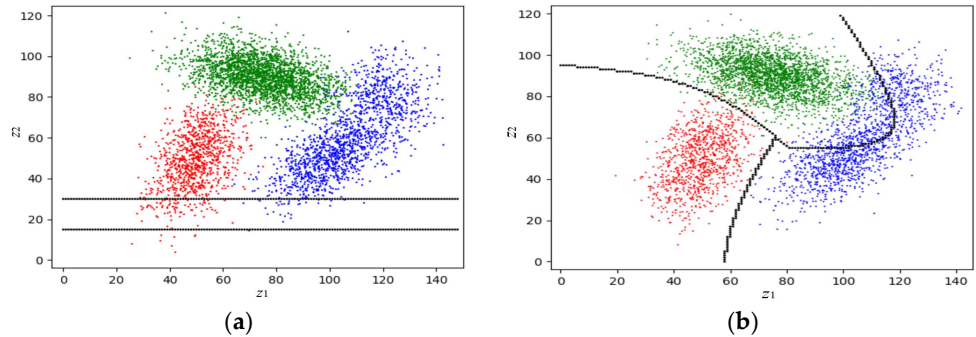

(a)  (b)

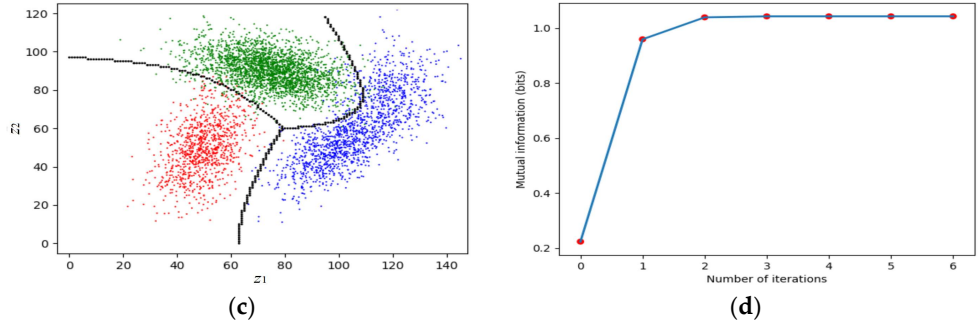

(c)        (d)

**Figure 6.** The MMI classification with a very bad initial partition. The convergence is very fast and stable without considering gradients. (**a**) The very bad initial partition. (**b**) The partition after the first iteration. (**c**) The partition after the second iteration. (**d**) The mutual information changes with iterations.

4.4.3. Explaining and Improving the EM Algorithm for Mixture Models

We know formula $P(x) = \sum_j P(y_j)P(x|y_j)$. For a given sampling distribution $P(x)$, we use the mixture model $P_\theta(x) = \sum_j P(y_j)P(x|\theta_j)$ to approximate to $P(x)$, making relative entropy $H(P\|P_\theta)$ close to 0. After setting the initial $P(x|\theta_j)$ and $P(y_j)$, $j = 1, 2, …$, we do the following iterations, each of which includes two matching steps (for details, see [13,73]):

**Matching 1:** Let the Shannon channel $P(y|x)$ match the semantic channel by repeating the following two formulas $n$ times:

$$P(y_j|x) = P(y_j)P(x|\theta_j)/P_\theta(x), \ P_\theta(x) = \sum_j P(y_j)P(x|\theta_j),$$
$$P^{+1}(y_j) = \sum_x P(x_i)P(y_j|x_i). \tag{44}$$

Of all the above steps, only the first step that changes $\theta_j$ increases or decreases $R$; other steps only reduce $R$.

**Matching 2**: Let the semantic channel match the Shannon channel to maximize $G$ by letting

$$P(x|\theta_j^{+1}) = P(x)P(x|\theta_j)/P_\theta(x), \ P_\theta(x) = \sum_j P(y_j)P(x_i|\theta_j). \tag{45}$$

End the iterations until $\theta$ or $H(P\|P_\theta)$ cannot be improved.
For the convergence proof, we can deduce [13]:

$$H(P\|P_\theta) = R - G + H(P^{+1}(y)\|P(y)). \tag{46}$$

This formula can also be used to explain pre-training in deep learning.

Since Matching 2 maximizes $G$, and Matching 1 minimizes $R$ and $H(P^{+1}(y)\|P(y))$, $H(P\|P_\theta)$ can approach 0.

The above algorithm can be called the EnM algorithm, in which we repeat Equation (44) $n$ or fewer times (such as $n \leq 3$) for $P^{+1}(y) \approx P(y)$. The EnM algorithm can perform better than the EM algorithm in most cases. Moreover, the convergence proof can help us avoid blind improvements.

Figure 7 shows an example of Gaussian mixture models for comparing the EM and E3M algorithms. The real model parameters are $(\mu_1, \mu_2, \sigma_1, \sigma_2, P(y_1)) = (100, 125, 10, 10, 0.7)$.

This example reveals that the EM algorithm can converge only because semantic or predictive mutual information $G$ and ShMI $R$ approach each other, not because the complete data's log-likelihood $Q$ continuously increases. The detailed discussion can be found in Section 3.3 of [73].

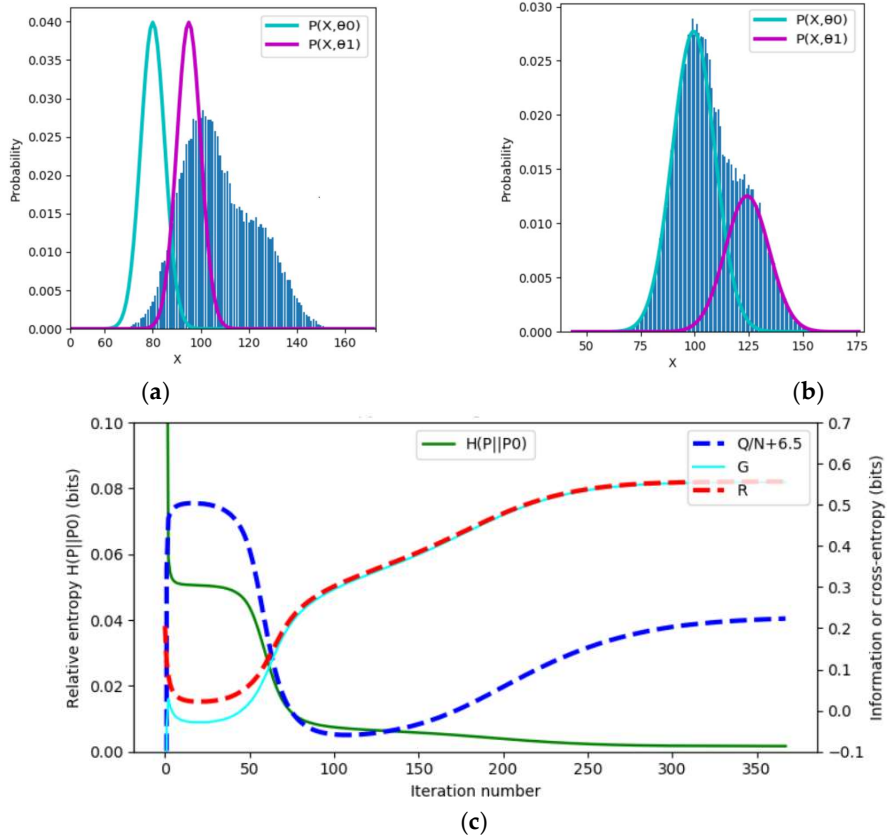

**Figure 7**. Comparing EM and E3M algorithms with an example that is hard to converge. The EM algorithm needs about 340 iterations, whereas the E3M algorithm needs about 240 iterations. In the convergent process, complete data log-likelihood $Q$ is not monotonously increasing. $H(P||P_\theta)$ decreases with $R − G$. (**a**) Initial components with $(\mu_1, \mu_2) = (80, 95)$. (**b**) Globally convergent two components. (**c**) $Q$, $R$, $G$, and $H(P||P_\theta)$ changes with iterations (initialization: $(\mu_1, \mu_2, \sigma_1, \sigma_2, P(y_1)) = (80, 95, 5, 5, 0.5)$).

Note that how many iterations are needed mainly depends on the initial parameters. For example, suppose we use initialization $(\mu_1, \mu_2, \sigma_1, \sigma_2, P(y_1)) = (105, 120, 5, 5, 0.5)$ according to the fair competition principle [73]. In that case, the EM algorithm needs about four iterations, whereas the E3M algorithm needs about three iterations, on average. Ref. [73] provides an initialization map, which can tell if a pair of initial means $(\mu_1, \mu_2)$ is good.

Refs. [13,73] provide more examples for testing the EnM algorithm.

## 5. Discussion 1: Clarifying Some Questions

*5.1. Is Mutual Information Maximization a Good Objective?*

Some researchers [3,24] actually maximize EMI, but they explain that we need to maximize ShMI. Since EMI is the lower limit of ShMI, we can reach maximum ShMI by maximizing EMI. However, Tschannent et al. doubt that ShMI maximization is a good objective [25]. The author thinks their doubt deserves attention.

From the perspective of the G theory, it is incorrect to say that EMI is the lower limit of ShMI because EMI changes with ShMI when we change the Shannon channel. If we optimize model parameters to maximize EMI, the Shannon channel and ShMI do not change. Generally, SeMI or EMI maximization is our purpose. In some cases, we need to maximize ShMI only because ShMI is the upper limit of SeMI. In the following, we discuss information maximization in different circumstances.

For supervised learning, such as multi-label learning ($X{\longrightarrow}Y$), in the learning stage, the sampling distribution $P(x, y)$ constrains ShMI, which is the upper limit of SeMI. This upper limit cannot be increased. However, in the classification stage, we classify $x$ according to truthlikeness functions $m_\theta(x, y_j)$, $j$ = 1, 2, … to maximize ShMI and SeMI simultaneously. According to the $R(G)$ function, the classification is to make $s{\longrightarrow}\infty$, so that $P(y_j|x)$ = 1 or 0 (see Equation (38)), and both $R$ and $G$ reach the highest point on the right side of the $R(G)$ curve (see Figure 4). When $s$ increases from 1, information efficiency $G/R$ decreases. Sometimes, we have to balance between maximizing $G$ and maximizing $G/R$.

For semi-supervised learning, such as the MaxMI classification of unseen instances (denoted as $X{\longrightarrow}Z{\longrightarrow}Y$, see Figure 5), SeMI is a reward function, and ShMI is its upper limit. We increase SeMI by maximizing ShMI, and, finally, we maximize both.

The following section discusses the maximization or minimization of $R$ and $G$ in deep learning.

Why is ShMI alone not suitable as an objective function in semantic communication? This is because it has nothing to do with right or wrong. For example, we classify people into four classes with labels: $y_1$ = "children", $y_1$ = "youth", $y_3$ = "adult", and $y_4$ = "elderly", according to their ages $x$. ShMI reaches its maximum when the classification makes $P(y_1)$ = $P(y_2)$ = $P(y_3)$ = $P(y_4)$ = 1/4. However, this is not our purpose. In addition, even if we exchange the labels of two age groups {children} and {elderly}, ShMI does not change. However, if labels are misused, the SeMI will be greatly reduced or negative. Therefore, it is problematic to maximize ShMI alone. However, SeMI maximization can ensure less distortion and more predictive information and is equivalent to - likelihood maximization and compatible with the RLS criterion. In addition, ShMI maximization is not a good objective because using Shannon's posterior entropy $H(Y|X)$ as the loss function is improper (see Section 5.4).

*5.2. Interpreting DNNs: The R(G) Function vs. the Information Bottleneck*

We take the AutoEncoder as an example to discuss the changes in ShMI and SeMI in self-supervised learning. The Autoencoder has a structure $X{\longrightarrow}Y{\longrightarrow}\hat{X}$, where $X$ is an input image, $Y$ is a low-dimensional representation of $X$, and $\hat{X}$ is the estimate of $X$. The successful learning method [69] is that first, we pre-train the network parameters of a multi-layer RBM from $X$ to $Y$, similar to solving a mixed model, to minimize the relative entropy $H(P||P_\theta)$. Then we decode $Y$ into $\hat{X}$ and fine-tune the network parameters between $X$ and $\hat{X}$ to minimize the loss with the RLS criterion, like maximizing EMI $I(X; \hat{X}_\theta)$.

From the perspective of the G theory, pre-training is to solve the mixed model so that $R = G$ between $X$ and $Y$, and the relative entropy is close to 0. Fine-tuning is to maximize $R$ and $G$. In terms of the $R(G)$ function (see Figure 4), pre-training is to find the point where $R = G$ and $s = 1$. Fine-tuning increases $R$ and $G$ so that $s = s_{max}$ by reducing the distortion between $X$ and $\hat{X}$.

Can we increase $s$ to improve $I(X; \hat{X}_\theta)$? Theoretically, it is feasible. With $T(\theta_j|x)$ = $\exp[-sd(x, y_j)]$ ($s > 0$), increasing $s$ is to narrow the coverage of the truth function (or the membership function of fuzzy set $\theta_j$). If both $x$ and $x_j$ are elements in the fuzzy set $\theta_j$, the average distortion between them will also be reduced. As for whether increasing $s$ is enough, it needs to be tested.

According to the information bottleneck theory [27], to optimize a DNN with structure $X{\longrightarrow}T_1{\longrightarrow}...{\longrightarrow}T_m{\longrightarrow}Y$, such as a DBN, we need to minimize $I(T_{i-1}; T_i) - \beta I(Y; T_{i-1}|T_i)$ ($i$ = 1, 2, …, $m$), like solving a $R(D)$ function. This idea is very inspiring. However, the author believes that every latent layer ($T_i$) needs its SeMI maximization and ShMI minimization. The reasons are:

- DNNs often need pre-training and fine-tuning. In the pre-training stage, the RBM is used for every latent layer.

- The RBM is like the EM algorithm for mixture models [78,79]. The author has proved (see Equation (46)) that we need to maximize SeMI and minimize ShMI to make mixture models converge.

On the other hand, the information bottleneck theory maximizes ShMI between $X$ and $\hat{X}$ or $Y$ and $\hat{Y}$. In contrast, the G theory suggests maximizing SeMI between $X$ and $\hat{X}$ or $Y$ and $\hat{Y}$.

*5.3. Understanding Gibbs Distributions, Partition Functions, MinMI Matching, and RBMs*

In statistical mechanics, the Gibbs distribution can be expressed as a distribution over different phase lattices or states:

$$P(e_k | T) = \exp(-\frac{e_k}{kT}) / Z, \ Z = \sum_l \exp(-\frac{e_l}{kT}). \tag{47}$$

It can also be expressed as a distribution over different energy levels [43]:

$$P(e_i | T) = P(e_i) \exp(-\frac{e_i}{kT}) / Z, \ Z = \sum_l P(e_l) \exp(-\frac{e_l}{kT}). \tag{48}$$

We can understand energy as distortion and $\exp[-e_i/(kT)]$ as the truth function. For machine learning, to predict the posterior probability of $x_i$, we have:

$$P(x_i | \theta_j) = P(x_i) T(\theta_j | x_i) / T(\theta_j), \ T(\theta_j) = \sum_k P(x_k) T(\theta_j | x_k) \tag{49}$$

which is similar to Equation (48).

Logical probability $T(\theta_j)$ is the partition function $Z$. If we put $T(\theta_j|x) = m(x, y_j)/mm_j$ into $T(\theta_j)$, there is

$$T(\theta_j) = \sum_i P(x_i) m(x_i, y_j) / mm_j = \sum_i P(x_i | y_j) / mm_j = 1 / mm_j. \tag{50}$$

From Equations (49) and (50), we derived $P(x|\theta_j) = P(x|y_j)$. We can see that the summation for $T(\theta_j)$ is only to get $T(\theta_j) = 1/mm_j$ so that $1/mm_j$ in the numerator and the denominator of the Gibbs distribution are eliminated simultaneously. Then the distribution $P(x|\theta_j)$ approximates to $P(x)m(x, y_j) = P(x|y_j)$. It does not matter how big $mm_j$ is.

Some researchers use the SoftMax function to express learning function $P(x, y|\theta_j)$ and the probability prediction, such as $P(x, y|\theta_j) = \exp[-d(x, y_j)]$ and

$$P(x | \theta_j) = \frac{P(x, y_j | \theta)}{\sum_k P(x_k, y_j | \theta)} = \frac{\exp[-d(x, y_j)]}{\sum_k \exp[-d(x_k, y_j)]}. \tag{51}$$

This expression is very similar to that of $m_\theta(x, y_j)$ in the semantic information method, but they are different. In the semantic information method,

$$P(x | \theta_j) = P(x) m_\theta(x, y_j) = \frac{P(x) \exp[-d(x, y_j)]}{\sum_k P(x_k) \exp[-d(x_k, y_j)]}, \tag{52}$$

in which there are $P(x)$ and $P(x_k)$.

Given $P(x)$ and semantic channel $T(y|x)$, we can order $P(y_j|x) = kT(y_j|x)$ for MinMI matching and then obtain $P(y_j) = kT(y_j)$, $k = 1/\sum_j T(y_j)$. Hence,

$$P(y_j) = T(y_j) / \sum_j T(y_j),$$
$$P(y_j | x) = T(y_j | x) / \sum_j T(y_j). \tag{53}$$

Note that if $T(y|x)$ also needs optimization, we can only use the MinMI iteration (see Section 6.2). If we use the above formula, $T(y|x)$ cannot be further optimized.

In addition, for a given $m(x, y)$, $P(x)$ and $P(y)$ are related; we cannot simply let $P(x, y) = P(x)m(x, y)P(y)$ because new $P(y)$ may not be normalized, that is,

$$P^{+1}(y_j) = \sum_i P(x_i)m(x_i, y_j)P(y_j), j = 1, 2, ...$$
$$\sum_j P^{+1}(y_j) \neq 1. \tag{54}$$

Nevertheless, we can fix one of $P(x)$ and $P(y)$ to get another. For example, given $P(x)$ and $m(x, y)$, we first obtain $T(y_j|x) = m(x, y_j)/mm_j$, then use Equation (53) to get $P(y)$.

Equation (54) can help us further understand the RBM. A RBM with structure $V \longrightarrow H$ [60] contains a set of parameters: $\theta = \{a_i, w_{ij}, b_j | i = 1, 2, ..., n; j = 1, 2, ..., m\}$. Parameters $a_i$, $w_{ij}$, and $b_j$ are associated to $P(v_i)$, $m_\theta(v_i, h_j)$, and $P(h_j)$, respectively. Optimizing $\{w_{ij}\}$ (weights) improves $m_\theta(v, h)$ and maximizes SeMI, and optimizing $\{b_j\}$ improves $P(h)$ and minimizes ShMI. Alternate optimization makes SeMI and ShMI close to each other.

*5.4. Understanding Fuzzy Entropy, Coverage Entropy, Distortion, and Loss*

We first assume that $\theta$ is a crisp set to see the properties of coverage entropy $H(Y_\theta)$.

We take age $x$ and the related classification label $y$ as an example to explain the coverage entropy.

Suppose the domain $U$ of $x$ is divided into four crisp subsets with boundaries: 15, 30, and 55. The labels are $y_1$ = "Children", $y_2$ = "Youth", $y_3$ = "Middle-aged people", and $y_4$ = "Elderly". The four subsets constitute a partition of $U$. We divide $U$ again into two subsets according to whether $x \geq 18$ and add labels $y_5$ = "Adult" and $y_6$ = "Non-adult". Then six subsets constitute the coverage of $U$.

The author has proved [16] that the coverage entropy $H(Y_\theta)$ represents the MinMI of decoding $y$ to generate $P(x)$ for given $P(y)$. The constraint condition is that $P(x|y_j) = 0$ for $x \notin \theta_j$. Some researchers call this MinMI the complexity distortion [77]. A simple proof method is to make use of the $R(D)$ function. We define the distortion function:

$$d(x, y_j) = \begin{cases} 0, & x \in \theta_j, \\ \infty, & x \notin \theta_j. \end{cases} \tag{55}$$

Then we obtain

$$R(D = 0) = H(Y_\theta) = -\sum_j P(y_j) \log T(\theta_j). \tag{56}$$

If these subsets form a partition of U, the coverage entropy will become the Shannon entropy.

If the above subsets are fuzzy, and the constraint condition is $P(x|y_j) \leq P(x|\theta_j)$ for $T(\theta_j|x) < 1$, then the minimum ShMI equals the coverage entropy minus the fuzzy entropy, that is, $R = I(X; Y) = H(Y_\theta) - H(Y_\theta|X)$ [18].

Shannon's distortion function also ascertains a semantic information quantity. Given the source $P(x)$ and the distortion function $d(x, y)$, we can get the truth function $T(\theta_j|x) = \exp[-d(x, y_j)]$ and the logical probability $T(\theta_j) = Z_j$. Then letting $P(y_j|x) = kT(\theta_j|x)$, we can get $P(y_j|x) = \exp[-d(x, y_j)]/\sum_j T(\theta_j)$. Furthermore, we have $I(X; Y) = I(X; Y_\theta) = H(Y_\theta) - \overline{d}$, which means that the minimum ShMI for a given distortion limit can be expressed as SeMI, and the SeMI decreases with the average distortion increasing.

Many researchers also use Shannon's posterior entropy $H(Y|X)$ as the average distortion (or loss) or $-\log P(y|x_i)$ as the distortion function. This usage shows a larger distortion when $P(y|x_i)$ is much less than 1. However, from the perspective of semantic communication, less $P(y|x_i)$ does not mean larger distortion. For example, with ages, when the above six labels are used on some occasions, $y_6$ = "Non-adult" is rarely used, so $P(y_6|x < 18)$ is very small, and hence $-\log P(y_6|x < 18)$ is very large. However, there is no distortion because $T(\theta_6|x < 18) = 1$. Therefore, Using $H(Y|X)$ to express distortion or loss is often unreasonable. In addition, it is easy to understand that for a given $x_i$, the distortion of a label has nothing to do with the frequency in which the label is selected, whereas $P(y|x_i)$ is related to $P(y)$.

Using $D_{KL}$ to represent distortion or loss [27] has a similar problem, whereas using semantic KL information to represent negative loss is appropriate.

Since SeMI or EMI can be written as the coverage entropy minus the average distortion, it can be used as a negative loss function.

*5.5. Evaluating Learning Methods: The Information Criterion or the Accuracy Criterion?*

In my opinion, there is an extraordinary phenomenon in the field of machine learning: researchers use the maximum likelihood, minimum cross-entropy, or maximum EMI criterion to optimize model parameters, but in the end, they prove that their methods are good because they have high classification accuracies for open data sets. Given this, why do they not use the maximum accuracy or minimum distortion criterion to optimize model parameters? Of course, one may explain that the maximum accuracy criterion cannot guarantee the robustness of the classifier; a classifier may be ideal for one source $P(x)$ but not suitable for another. However, can the robustness of classifiers be fairly tested with open data sets?

The author believes that using information criteria (including the maximum likelihood criterion) is not only for improving robustness but also for reducing the underreporting of small probability events or increasing the detection rate of small probability events. The reason is that the information loss due to failing to report smaller probability events is greater in general; the economic loss is also greater in general, such as in medical tests, X-ray image recognitions, weather forecasts, earthquake predictions, etc.

How do we resolve the inconsistency between learning and evaluation criteria? Using information criteria to evaluate various methods is best, but it is not feasible in practice. Therefore, the author suggests that in cases where overall accuracies are not high, we also check the accuracy of small probability events in addition to the overall accuracy. Then we use the vector composed of two accuracies to evaluate different learning methods.

## 6. Discussion2: Exploring New Methods for Machine Learning

*6.1. Optimizing Gaussian Semantic Channels with Shannon's Channels*

It is difficult to train the truth function expressed as SoftMax functions with samples or sampling distributions, where Gradient Descent is usually used. Suppose we can convert SoftMax function learning into Gaussian function learning without considering gradients. In that case, the learning will be as simple as the EM algorithm of the Gaussian mixture models.

Suppose that $P(y_j|x)$ is proportional to a Gaussian function. Then, $P(y_j|x)/\sum_k P(y_j|x_k)$ is a Gaussian distribution. We can assume $P(x) = 1/|U|$ ($|U|$ is the number of elements in $U$) and then use $P(y_j|x)$ to optimize $T(\theta_j|x)$ by maximizing semantic KL information:

$$I(X;\theta_j) = \sum_i P(x_i|y_j) \log \frac{T(\theta_j|x_i)}{T(\theta_j)} = \sum_i \frac{P(y_j|x_i)}{\sum_k P(y_j|x_k)} \log \frac{T(\theta_j|x_i)}{\sum_k T(\theta_j|x_k)} + \log|U|. \quad (57)$$

Obviously, $I(X; \theta_j)$ reaches its maximum as $T(\theta_j|x) \propto P(y_j|x)$, which means that we can use the expectation $\mu_j$ and the standard deviation $\sigma_j$ of $P(y_j|x)$ as those of $T(\theta_j|x)$.

In addition, if we only know $P(x|y_j)$ and $P(x)$, we can replace $P(y_j|x)$ with $m(x, y_j) = P(x|y_j)/P(x)$, and then optimize the Gaussian truth function. However, this method requires that no $x$ makes $P(x) = 0$. For this reason, we need to replace $P(x)$ with an uninterrupted distribution close to $P(x)$.

The above method of optimizing Gaussian truth functions does not need Gradient Descent.

*6.2. The Gaussian Channel Mixture Model and the Channel Mixture Model Machine*

So far, only the mixture model that consists of likelihood functions, such as Gaussian likelihood functions, has been used. However, we can also use Gaussian truth functions to construct the Gaussian Channel Mixture Model (GCMM).

We can also use the EnM algorithm for the GCMM. **Matching 1** and **Matching 2** becomes:

**Matching 1:** Let the Shannon channel match the semantic channel by using $P(x|\theta_j)$ = $P(x)T(\theta_j|x)/T(\theta_j)$ and repeating Equation (44) $n$ times.

**Matching 2**: Let the semantic channel match the Shannon channel by letting:

$$T(\theta_j^{+1}|x) = \exp\left[\frac{-(x-\mu_j)^2}{2\sigma_j^2}\right] \propto \frac{P(y_j|x)}{P(y_j)} = \frac{P(x)T(\theta_j|x)}{T(\theta_j)P_\theta(x)} \propto \frac{P(x)T(\theta_j|x)}{P_\theta(x)}, \quad (58)$$

which means we can use the expectation and the standard deviation of $P(y_j|x)$ or $P(x)T(\theta_j|x)/P_\theta(x)$ as those of $T(\theta_j^{+1}|x)$.

We can set up a neural network working as a channel mixture model and call it the Channel Mixture Model Machine (CMMM). Figure 8 includes a typical neuron and a neuron in the CMMM. In Figure 8b, the truth value $T(\theta_j|x_i)$ is used as weight $w_{ji}$. The input may be $x_i$, a vector **x**, or a distribution $P(\mathbf{x})$, for which we use different methods to obtain the same $T(\theta_j)$.

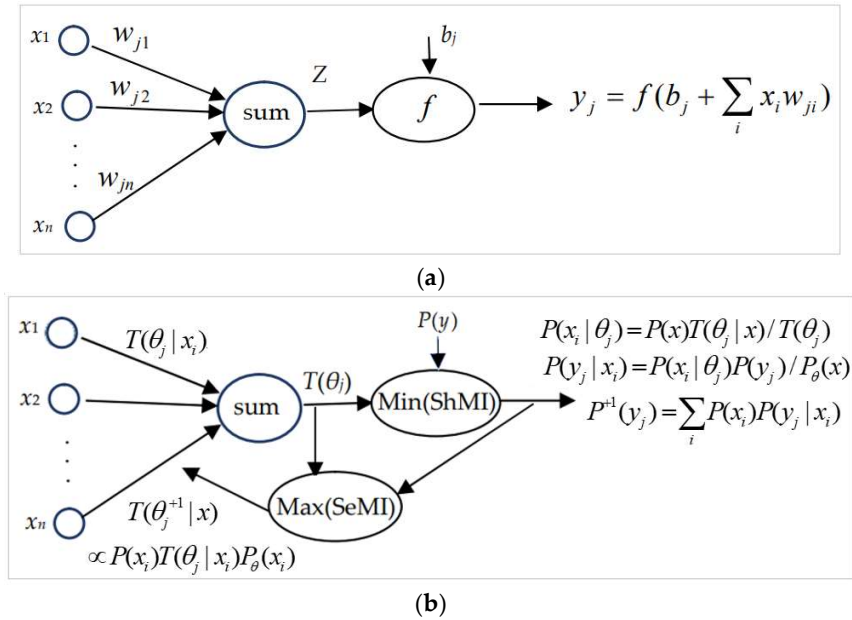

**Figure 8.** Comparing a typical neuron and a neuron in a CMMM. (**a**) A typical neuron in neural networks. (**b**) A neuron in the CMMM and its optimization.

As shown in Figure 8a, $f$ is an activation function in a typical neuron. If $b_j = P(y_j) - T(\theta_j) \leq 0$ and $f$ is a Relu function, two neurons will be equivalent.

We only need to repeatedly assign values to optimize the model parameters in the CMMM. The initial truth values or weights may be between 0 and a positive number $c$ because replacing $T(\theta_j|x)$ with $cT(\theta_j|x)$ does not change $P(x|\theta_j)$.

The author's preliminary experiments show that a CMMM works as expected. The convergence is very fast. Optimizing weights only needs 3–4 iterations. We need more experiments to test whether we can use CMMMs or Gaussian CMMMs (weight distributions are limited to Gaussian functions) to replace RBMs.

*6.3. Calculating the Similarity between any Two Words with Sampling Distributions*

Considering the similarity between two words, $x$ and $y$, we can get the similarity from a sampling distribution:

$$S(x,y) = \begin{cases} 0, & \text{if } P(x,y) = 0, \\ \dfrac{m(x,y)}{\max[m(x,y)]}, & \text{otherwise,} \end{cases} \quad (59)$$

where $\max[m(x, y)]$ is the maximum of all $m(x, y)$. The similarity obtained in this way is symmetrical. This similarity can be regarded as relative truthlikeness. The advantages of this method are:

- It is simple without needing the semantic structure such as that in WordNet;
- This similarity is similar to the improved PMI similarity [65] which varies between 0 and 1;
- This similarity function is suitable for probability predictions.

The disadvantage of this method is that the semantic meaning of the similarity function may be unclear. We can limit $x$ and $y$ in different and smaller domains and get asymmetrical truth functions (see Equation (34)) that can better indicate semantic meanings.

Some semantic similarity measures, such as Resnik's similarity measure [23]), based on the structure of WordNet, reflect philosophical thought that a hypothesis with less logical probability can carry more information content. They are similar to truthlikeness $m_\theta(x, y)$ and suitable for classification, whereas $S(x, y_j)$ defined above is suitable for probability predictions. We can convert $S(x, y_j)$ with $P(x)$ into the truthlikeness function for classification.

Latent Semantic Analysis (LSA) has been successfully used in natural language processing and the Transformer. How to combine the semantic information method with LSA to get better results is worth exploring.

*6.4. Purposive Information and the Information Value as Reward Functions for Reinforcement Learning*

The G measure can be used to measure purposive information (about how the result accords with the purpose) [80]. For example, a passenger and a driver in a car get information from a GPS map about how far they are from their destination. For the passenger, it is purposive information, while for the driver, it is the feedback information for control. If we train the machine to drive, the purposive information can be used as a reward function for reinforcement learning.

Generally, control systems use error as the feedback signal, but Wiener proposes using entropy to indicate the uncertainty of control results. Sometimes we need both accuracy and precision. The G measure meets this purpose.

Researchers have used Shannon information theory for constraint control and reinforcement learning [81,82]. ShMI (or $D_{KL}$) represents necessary control complexity. Given distortion $D$, the smaller the mutual information $R$, the higher the control efficiency. When we use the $G$ theory for constraint control and reinforcement learning, the SeMI measure can be used as the reward function, then $G/R$ represents the control efficiency. We explain reinforcement learning as being like driving a car to a destination. We need to consider:

- Choosing an action $a$ to reach the destination;
- Learning the system state's change from $P(x)$ to $P(x|a)$;
- Setting the reward function, which is a function of the goal, $P(x)$, and $P(x|a)$.

Some problems that reinforcement learning needs to resolve are:

1. How to get $P(x|a)$ or $P(x|a, h)$ ($h$ means the history)?
2. How to choose an action $a$ according to the system state and the reward function?
3. How to achieve the goal economically, that is, to balance the reward maximization and the control-efficiency maximization?

In the following, we suppose that $P(x|a)$ is known already; only problems 2 and 3 need to be resolved.

Sometimes, the goal is not a point but a fuzzy range for the control of uncertain events. For example, a country, whose average death age is 50, sets a goal: $y_j$ = "The death ages of the population had better not be less than 60 years old". Suppose that the control method is to improve medical conditions. In this case, Shannon information represents the control cost, and the semantic information indicates the control effect. Although we can raise the average death age to 80 at a higher cost, it is uneconomical in terms of control efficiency. There is enough purposive information when the average death age reaches about 62. The author in [80] discusses this example.

Assume that the goal is expressed as Logistic function $T(\theta_j|x) = 1/[1 + \exp(-0.8(x - 60))]$ (see Figure 9). The prior distribution $P(x)$ is normal ($\mu$ = 50 and $\sigma$ = 10), and the control result $P(x|a_j)$ of a medical condition $a_j$ is also normal. The purposive information or the reward function of $a_j$ is:

$$I(X;\theta_j|a_j) = \sum_i P(x_i|a_j)\log\frac{T(\theta_j|x_i)}{T(\theta_j)} = \sum_i P(x_i|a_j)\log\frac{P(x_i|\theta_j)}{P(x_i)}. \quad (60)$$

We can let $P(x|a_j)$ approximate to $P(x|\theta_j)$ by changing the $\mu_j$ and $\sigma_j$ of $P(x|a_j)$ to get the minimum KL information $I(X; a_j)$ so that information efficiency $I(X; \theta_j|a_j)/I(X; a_j)$ is close to 1. In addition, we can increase $s$ in the following formula to increase both SeMI and ShMI:

$$P(x_i|a_j) \approx P(x|\theta_j|s) = P(x_i)T(\theta_j|x_i)^s \Big/ \sum_k P(x_k)T(\theta_j|x_i)^s. \quad (61)$$

The information efficiency will decrease with $s$ increasing from 1. So, we need the trade-off between information efficiency maximization and purposive information maximization.

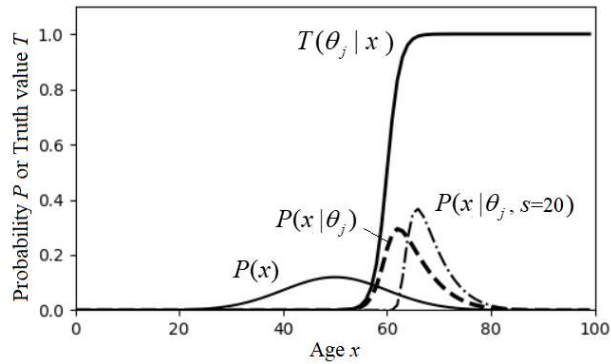

**Figure 9.** Illustrating population death age control for measuring purposive information. $P(x|a_j)$ approximates to $P(x|\theta_j) = P(x|\theta_j, s = 1)$ for information efficiency $G/R = 1$. $G$ and $R$ are close to their maxima as $P(x|a_j)$ approximates to $P(x|\theta_j, s = 20)$.

The author, in [80], provides the computing results for this example. We assume $Y = y_j$, then SeMI equals $I(X; \theta_j|a_j)$, and ShMI equals $I(X; a_j)$. The results indicate that the highest information efficiency $G/R$ is 0.95 when $P(x|a_j)$ approximates to $P(x|\theta_j)$. When $s$ increases from 1 to 20, the purposive information $G$ increases from 2.08 bits to 3.13 bits, but the information efficiency $G/R$ decreases from 0.95 to 0.8. To balance $G$ and $G/R$, we should select $s$ between 5 and 15.

If there are multiple goals, we must optimize $P(a_j) = P(y_j)$, $j = 1, 2, \ldots$ according to Equation (38).

Reinforcement learning is also related to information value.

Cover and Thomas have employed the information-theoretic method to optimize the portfolio [83]. The author has used semantic information to optimize the portfolio [84]. The author obtained the value-added entropy formula and the information value formula (see Appendix B for details):

$$H_v(\mathbf{R}, \mathbf{q} | \theta_j) = \sum_i P(\mathbf{x}_i | \theta_j) \log R_i(\mathbf{q}), \tag{62}$$

$$V(\mathbf{X}; \theta_j) = \sum_i P(\mathbf{x}_i | \theta_j) \log(R_i(\mathbf{q}^{**}) / R_i(\mathbf{q}^*)). \tag{63}$$

$H_v(\mathbf{R}, \mathbf{q}|\theta_j)$ is the value-added entropy, where $\mathbf{q}$ is the vector of portfolio ratios, $R_i(\mathbf{q})$ is the return of the portfolio with $\mathbf{q}$ when the $i$-th price vector $\mathbf{x}_i$ appears, and $\mathbf{R}$ is the return vector. $V(\mathbf{X}; \theta_j)$ is the predictive information value, where $\mathbf{X}$ is a random variable taking $\mathbf{x}$ as its value, $\mathbf{q}^*$ is the optimized vector of portfolio ratios according to the prior distribution $P(\mathbf{x})$, and $\mathbf{q}^{**}$ is that according to prediction $P(\mathbf{x}|\theta_j)$. $V(\mathbf{X}; \theta_j)$ can be used as a reward function to optimize probability predictions and decisions. However, to calculate the actual information value (in the learning stage), we need to replace $P(\mathbf{x}|\theta_j)$ with $P(\mathbf{x}|y_j)$ in Equation (63).

From the value-added entropy formula, we can derive some useful formulas. For example, if an investment will result in two possible results, yield rate $r_1 < 0$ with probability $P_1$ and yield rate $r_2 > 0$ with probability $P_2$, the optimized investment ratio is

$$q^* = E/|r_1 r_2|, \tag{64}$$

where $E$ is the expected yield rate. If $r_1 = -1$, we have $q^* = P_2 - P_1/r_2$, which is the famous Kelly formula. For the above formula, we assume that the risk-free rate $r_0 = 0$; otherwise, there is

$$q^* = E_0 R_0 / |r_{10} r_{20}|, \tag{65}$$

where $E_0 = E - r_0$, $r_{10} = r_1 - r_0$, $r_{20} = r_2 - r_0$.

However, $V(\mathbf{X}; \theta_j)$ is only the information value in particular cases. Information values in other situations need to be further explored.

*6.5. The Limitations of the Semantic Information G Theory*

Based on the G theory, we seem to be able to apply the channels matching algorithms (for muti-label learning, MMI classifications, and mixture models) to DNNs for various learning tasks. However, DNNs include not only statistical reasoning but also logical reasoning. The latter is realized by setting biases and activation functions. For example, the Relu function is the fuzzy logic operation, which the author uses in establishing the decoding model of color vision [54,55], whereas the semantic information-theoretic method now only involves the statistical part. It helps explain deep learning, but it is far from enough to guide one in building efficient DNNs.

DNNs have shown strong vitality in many aspects, such as feature extraction, self-supervised learning, and reinforcement learning. However, the G theory is not sufficient as a complete semantic information theory. Instead, it should be combined with feature extraction and fuzzy reasoning for further developments to keep pace with deep learning.

**7. Conclusions**

Looking back at the evolutionary histories of semantic information measures and learning functions, we can find that two parallel trends lead to the same goal. In the end, both use the truth or similarity function and the maximum semantic or estimated information criterion to optimize semantic communication and machine learning. The estimated information measure is a special case of the semantic information measure. The author proposed both 30 years ago. The similarity function for deep learning is also the confusion probability function used in the semantic information G theory, and it is a particular truth function. Both are proportional to $m(x, y_j) = P(x, y_j)/[P(x)P(y_j)]$, and their maximum is one. Compared with the likelihood function and the anti-probability function (often expressed as the Logistic function), the similarity function is independent of prior probability distributions $P(x)$ and $P(y)$, has good transferability, and is more suitable for multi-label learning.

Using the G theory to analyze machine learning, we have concluded that Shannon mutual information $R$ is the upper limit of semantic mutual information $G$; $G/R$ indicates the information efficiency. Therefore, maximizing $G$ instead of $R$ is essential. $R$ sometimes needs to be maximized (as the upper limit of $G$) and sometimes minimized (to improve the information efficiency).

This paper has defined and discussed some essential functions (such as $m(x, y)$ and $m_\theta(x, y)$) and two generalized entropies (fuzzy entropy $H(Y_\theta|X)$ and coverage entropy $H(Y_\theta)$). Moreover, it has interpreted the Gibbs distribution as the semantic Bayes' prediction, the partition function as the logical probability, and the energy function as the distortion function.

This paper has clarified that in most cases, SeMI maximization rather than ShMI maximization is a good objective. It has used the $R(G)$ function to explain Autoencoders and similar learning tasks: pre-training makes Shannon's mutual information and semantic mutual information equal, that is, $G = R$ ($s = 1$); fine-tuning increases both $R$ and $G$ ($s{-}{>}\infty$). Unlike the information bottleneck theory, this paper has interpreted that RBMs used for pre-training also minimize the ShMI of every latent layer of DNNs.

The G theory and the evolution of semantic information measures and learning functions remind us that there are two potential opportunities. One is that we can use GCMMs or CMMMs to pre-train the latent layers of DNNs without needing Gradient Descent and Backpropagation. In the fine-tuning stage, we can increase $s$ in the $R(G)$ function to maximize both SeMI and ShMI. This method should be able to simplify and speed up deep learning distinctly. Another potential opportunity is to use the G measure as the reward function to optimize reinforcement learning. The G measure can help us balance between the purposive information and the information efficiency of control. In the future, topics related to the two opportunities are especially worth further studying.

However, deep learning also involves feature extraction and fuzzy reasoning. The G theory is far behind the requirements of deep learning. Therefore, it is a challenge to combine the G theory with deep learning for practical DNNs. Nevertheless, the G theory is expected to be the foundational part of a unified semantic information theory in the future. The semantic information theory and deep learning should be able to promote each other, making DNNs more interpretable.

**Acknowledgments:** The author thanks Peizhuang Wang for his long-term support and encouragement. The four Reviewers' comments helped the author significantly improve and enrich this paper. The author thanks them for their patience, comments, and lenient ratings. The author particularly thanks Viacheslav Kovtun. Without his encouragement, the author would not have finished this review.

**Conflicts of Interest:** The author declares no conflict of interest.

**Abbreviations**

BOYL              Bootstrap Your Own Latent

| DIM | Deep InfoMax (Information Maximization) |
| DNN | Deep Neural Network |
| DT | DeLuca-Termini |
| DV | Donsker-Varadhan |
| EM | Expectation-Maximization |
| EMI | Estimated Mutual Information |
| EnM | Expectation-n-Maximization |
| GCMM | Gaussian Channel Mixture Model |
| CMMM | Channel Mixture Model Machine |
| GPS | Global Positioning System |
| G theory | Semantic information G theory (G means generalization) |
| InfoNCE | Information Noise Contrast Estimation |
| ITL | Information-Theoretic Learning |
| KL | Kullback–Leibler |
| LSA | Latent Semantic Analysis |
| MaxMI | Maximum Mutual Information |
| MinMI | Minimum Mutual Information |
| MINE | Mutual Information Neural Estimation |
| MoCo | Momentum Contrast |
| PMI | Pointwise Mutual Information |
| SeMI | Semantic Mutual Information |
| ShMI | Shannon's Mutual Information |
| SimCLR | A simple framework for contrastive learning of visual representations |

**Appendix A. About Formal Semantic Meaning**

The semantic meaning of a label or concept contains both its denotation and connotation. The connotation includes various attributes. However, the formal semantic meaning only contains the denotation and one related attribute. For example, the connotation of "elderly" contains attributes: high age, doddery body, no working, etc. However, the formal semantic meaning of "elderly" only includes some ages covered by the denotation.

**Appendix B. The Definitions of the Value-Added Entropy and the Information Value**

Let variable $x$ be the price vector of a group of securities, and its constant is $\mathbf{x}_i = (x_{i1}, x_{i2}, \ldots)$, $i = 1, 2, \ldots$ The current price vector is $\mathbf{x}_0 = (x_{01}, x_{02}, \ldots)$, and the return vector is $\mathbf{R}_i = (R_0, x_{i1}/x_{01}, x_{i2}/x_{02}, \ldots)$, where $R_0 = 1 + r_0$ and $r_0$ is the risk-free rate. The vector of portfolio ratios is $\mathbf{q} = (q_0, q_1, q_2, \ldots)$. The return on investment is a function of $\mathbf{q}$, namely:

$$R_i(\mathbf{q}) = \sum_k q_k r_{ik} \tag{A1}$$

Assuming that the probability prediction is $P(\mathbf{x}|\theta_j)$, the *Value-added entropy* (that is, the expected doubling rate if the log is log$_2$) is:

$$H_v(\mathbf{R}, \mathbf{q} | \theta_j) = \sum_i P(\mathbf{x}_i | \theta_j) \log R_i(\mathbf{q}) \tag{A2}$$

where $\mathbf{q}$ represents the decision; it is also a learning function. We can increase the value-added entropy by optimizing $\mathbf{q}$. The predictive information value is

$$V(\mathbf{X}; \theta_j) = \sum_i P(x_i | \theta_j) \log[R_i(\mathbf{q}^{**}) / R_i(\mathbf{q}^*)] \tag{A3}$$

where $\mathbf{q}^*$ is the optimized vector of portfolio ratios according to the prior distribution $P(\mathbf{x})$, and $\mathbf{q}^{**}$ is that according to prediction $P(\mathbf{x}|\theta_j)$. The predictive information value can be used as a reward function to optimize probability predictions and decisions. How to calculate the actual information value (in the learning stage), we need to replace $P(\mathbf{x}|\theta_j)$ with $P(\mathbf{x}|y_j)$.